%% file: main.tex
\documentclass[reprint,aps,pra]{revtex4-2}
\usepackage{hyperref}
\usepackage[utf8]{inputenc}
\usepackage{graphicx}
\usepackage{physics}
\usepackage{bm}
\usepackage{xcolor}
\usepackage[caption=false]{subfig}
\usepackage{tikz}
\usepackage{cleveref}
\usepackage{orcidlink}
\usepackage{quantikz}
\usepackage{geometry}
\usepackage{MnSymbol} 
\usepackage{import}
\usepackage{acro}

\DeclareAcronym{DLR}{
short= DLR, 
long = German Aerospace Center
}

\DeclareAcronym{QEC}{
short= QEC, 
long = quantum error correction
}
\DeclareAcronym{QAOA}{
short = QAOA,
long = Quantum Approximate Optimization Algorithm
}
\DeclareAcronym{FGA}{
short = FGA,
long = flight-gate assignment
}
\DeclareAcronym{CNOT}{
short = CNOT,
long = controlled-not
}
\DeclareAcronym{QUBO}{
short = QUBO,
long = quadratic unconstrained binary optimization
}
\DeclareAcronym{NP}{
short = NP,
long = nondeterministic polynomial time
}
\DeclareAcronym{QPC}{
short = QPC,
long = quantum parity code
}
\DeclareAcronym{CSS}{
short = CSS,
long = Calderbank-Shor-Steane
}
\begin{document}
\title{Asymmetric quantum error correction efficiently tackles application-specific noise effects} 
\author{Abhishek Yadav\orcidlink{0009-0008-1124-7822}}
\email{Abhishek.Yadav@student.uni-siegen.de}
\author{Peter K. Schuhmacher\orcidlink{0000-0003-1232-4363}}
\email{peter.schuhmacher@dlr.de}
\author{Michael Epping\orcidlink{0000-0003-0950-6801}}
\email{michael.epping@dlr.de}
\affiliation{Institute of Software Technology, Department of High-performance Computing, German Aerospace Center (DLR), Sankt Augustin, Germany}
\date{\today}

\begin{abstract}
Noise is a major challenge for current quantum computers.
It can be broadly categorized into bit-flip and phase-flip errors.
These two types do not necessarily affect the executed algorithm, thus also the application, in the same way. 
We illustrate this general effect for the example of the \ac{QAOA} applied to a small instance of the \ac{FGA} problem. 
We compare bit-flip and phase-flip Pauli noise under both layer-level and gate-level noise models, using two circuit decompositions of the same ideal \ac{QAOA} unitary: a CNOT-based decomposition and a native-$R_{ZZ}$  decomposition.
For the six-qubit FGA instance considered here,
bit-flip noise produces a larger degradation in the
optimized QAOA performance across the investigated
circuit depths and noise models. The asymmetry is
most pronounced in the layer-level and native-$R_{ZZ}$
simulations. We explain this behavior through the
commutation and propagation properties of the errors
and their effects on the final computational-basis
measurements. Since these mechanisms are not specific
to the considered FGA instance, we expect similar
noise asymmetries to arise in other Ising-based
QAOA applications, although this remains to be
investigated. We then exploit these insights to tackle noise particularly efficiently using asymmetric error-correcting codes.
As an illustration, we use the \ac{QPC}, a generalization of the 9-qubit Shor code, and show that a smaller asymmetric code can achieve nearly the same improvement as a larger symmetric choice.
This suggests that allocating error-correction resources
according to algorithmic noise sensitivity, rather
than physical error rates alone, can be beneficial
even when the physical noise is symmetric.
Finally, we discuss how information about the noise obtained through calibration can be exploited in our approach.
\end{abstract}

\maketitle
\acresetall

 	\section{Introduction}
 	Currently available quantum devices can execute non-trivial quantum circuits, but their performance is still limited by noise, decoherence, and finite circuit depth \cite{Preskill2018quantumcomputingin}. 
 	It is thus crucial to handle the noise such that larger algorithms can be executed which are important for real-world problems.
 	Besides efforts to improve the hardware, there are essentially two approaches to tackle this problem: error mitigation and \ac{QEC}.
 	The former requires less resources, but does not scale to large computations \cite{Cai2023}.
 	The latter promises fault-tolerance in the future but comes with a significant resource overhead~\cite{Lidar_Brun_2013}. 
 	Most \ac{QEC} codes can correct as many bit-flip errors as phase-flip errors, but there are also explicitly asymmetric codes~\cite{ioffe2007asymmetric,roffe2023bias}, which may be more efficient, e.g. if the noise is not symmetric.
 	However, the advantage of \ac{QEC} has already been demonstrated~\cite{Ofek2016,Egan2021,Postler2022,Acharya2025,Bluvstein2026,lavasani2026quantumerrorcorrectioncolor}.
 	As \ac{QEC} starts to be beneficial, we are transitioning to the early fault-tolerant era.
 	Now it is very interesting to investigate how small \ac{QEC} codes can improve the performance of algorithms~\cite{Katabarwa2024,Dangwal2025}.
 	
 	A well-studied class of algorithms that fit on current quantum computers are so-called variational quantum algorithms~\cite{cerezo2021variationalreview}.
 	They are based on parametrized quantum circuits. 
 	For a fixed set of parameters, the circuit is executed repeatedly to estimate an objective function, often written as the expectation value of a problem-dependent observable. 
 	A classical optimizer then updates the parameters for the next circuit execution. 
 	In this way, the quantum device is used to generate and measure parametrized quantum states, while the classical computer performs the outer optimization loop.
 	
 	Among variational quantum algorithms, the \ac{QAOA} is one of the most widely studied examples. 
 	It was introduced by Farhi, Goldstone, and Gutmann for obtaining approximate solutions to combinatorial optimization problems in the gate model of quantum computation \cite{farhi2014quantumapproximateoptimizationalgorithm}. 
 	The algorithm is motivated by adiabatic quantum optimization, but \ac{QAOA} implements it as a finite-depth parametrized circuit rather than a continuous adiabatic process \cite{farhi2000quantumcomputationadiabatic,farhi2014quantumapproximateoptimizationalgorithm,Lidar_2018}.
 	Since its introduction, \ac{QAOA} has become one of the central variational algorithms for studying quantum optimization. 
 	For other approaches in quantum optimization we refer to the reviews \cite{Yarkoni_2022,au-yeung2023,Abbas2024}.
 	In \ac{QAOA}, the objective function is encoded into a problem Hamiltonian, whose low-energy states correspond to good candidate solutions and whose ground state encodes an optimal solution of the problem. 
 	Starting from an easy-to-prepare initial state, the algorithm applies alternating unitary evolutions generated by the problem Hamiltonian and by a non-commuting mixing Hamiltonian. 
 	The corresponding evolution angles are treated as variational parameters and are optimized by the classical optimizer. 
 	
 	\ac{QAOA} has been applied to several combinatorial optimization problems, including MaxCut~\cite{Maxcut_qaoa}, satisfiability problems~\cite{mas-ksat_qaoa}, graph coloring~\cite{graphcolouring_qaoa}, scheduling problems~\cite{Werner2026}, and more general \ac{QUBO} formulations \cite{blekos2024review}. 
 	In particular operational research and logistics is a field in which there are many hard combinatorial optimization problems to solve, and which could therefore benefit greatly from quantum optimization \cite{Bayerstadler2021}.
 	Of particular importance to this study is previous work on the flight gate assignment problem~\cite{stollenwerk2019flightgate,Chai_2023,quark2024}, which we introduce in detail below.
 	
 	Theoretical and numerical studies have investigated the approximation quality of \ac{QAOA}, its dependence on the circuit depth, and strategies for improving parameter initialization and optimization \cite{farhi2014quantumapproximateoptimizationalgorithm,zhou2020qaoa,Misra2023}. 
 	In parallel, experimental demonstrations have implemented \ac{QAOA} or \ac{QAOA}-inspired optimization on superconducting processors and trapped-ion quantum simulators~ \cite{Harrigan_2021,pagano2020qaoa}. 
 	These studies show that \ac{QAOA} is a useful testbed for understanding how variational quantum circuits behave under realistic hardware constraints.
 	On the one hand, new evidence for an advantage of \ac{QAOA} over the best-known classical algorithms has recently been found on the low autocorrelation binary sequences (LABS) problem~\cite{Shaydulin2024}, for Max-Cut and Max-Independent-Set~\cite{farhi2026lowerboundingmaxcuthigh}, for the Max-k-Cut problem \cite{apte2026provableclassicalquantumlocal}, and several combinatorial optimization problems when replacing the classical optimization by a fixed linear ramp~\cite{Montanez-Barrera2025}.
 	On the other hand progress on quantum-inspired classical algorithms like the mean-field approximate optimization algorithm has identified cases where \ac{QAOA} may not yield a quantum advantage, including Max-k-XOR~\cite{Mueller2025}.
 	But the study of these classical algorithms can also help to identify problems with potential for a quantum advantage, e.g. in the presence of an overlap gap property~\cite{goh2026overlapgappropertyseparating}.
 	We also mention that research on \ac{QAOA} continues to benefit from insights from many-body physics, see e.g. \cite{boulebnane2026decouplingqaoaindependentspinboson}.
 	
 	The examples above show that \ac{QAOA} is an interesting area of research.
 	However, the performance of \ac{QAOA} on current devices cannot be understood from the ideal circuit alone. 
 	Several studies have shown that noise can degrade the quality of the prepared state and modify the effective optimization landscape. 
 	Xue et al.~\cite{xue2019effects} studied \ac{QAOA} under typical noise channels, including dephasing, bit-flip, and depolarizing noise,  and observed that  output-state fidelity decreases with increase in the noise strength and the number of gates, leading to smaller gradients and a flatter parameter space.
 	Marshall et al.~\cite{Marshall_2020} further analyzed local noise in \ac{QAOA} circuits and derived approximate expressions for the fidelity and cost in terms of the noise rate, system size, and circuit depth, thereby making explicit the trade-off between deeper circuits and noise accumulation. 
 	Realistic hardware-oriented studies have shown that increasing the \ac{QAOA} depth does not always improve the observed performance on real devices. Although deeper circuits provide higher output fidelity in the ideal noiseless setting, they also require more physical gates and are therefore more affected by gate errors, decoherence, and connectivity-induced compilation overhead \cite{alam2019analysis,Harrigan_2021}. 
 	
 	Most noisy-\ac{QAOA} studies consider either symmetric depolarizing noise, standard dephasing, or separate comparisons of common noise channels. 
 	For many physical platforms, however, noise is not symmetric in the Pauli basis: bit-flip and phase-flip errors can occur with different probabilities.
 	This distinction is important because \ac{QAOA} alternates between Hamiltonians with different Pauli structure, so $X$-type and $Z$-type errors do not affect the optimization process in the same way. 
 	Therefore, as we will see below, a systematic comparison of bit-flip and phase-flip noise provides a basis for deciding whether an asymmetric quantum error-correction code can improve the noisy algorithmic output more efficiently than a symmetric one. 
 	There are thus three areas that have been fairly well studied: the effect of symmetric noise on \ac{QAOA}, asymmetric error-correcting codes, and the application of \ac{QAOA} to problems of industrial interest. 
 	The intersection between these topics remains largely unexplored. 
 	
 	In this article, we present a puzzle piece that bridges this gap:
 	A case-study in which we apply asymmetric \ac{QEC} to a variational quantum algorithm for a real-world optimization problem, which is not affected by bit-flip and phase-flip errors equally.
 	More specifically, we apply \ac{QAOA} to the \ac{FGA} problem~\cite{stollenwerk2019flightgate,Chai_2023,quark2024}. 
 	The \ac{FGA} problem is critical for efficient airport operations, where flights must be assigned to available gates in a way that minimizes overall  commuting time of the passengers at the airport. 
 	The problem is inherently combinatorial and maps naturally to a \ac{QUBO} formulation, making it an ideal candidate for \ac{QAOA}. 
 	
 	The aim of this work is to characterize how asymmetric Pauli noise affects \ac{QAOA} when the same problem Hamiltonian is simulated under different circuit and noise-model descriptions. 
 	We implement the \ac{QAOA} ansatz using two universal gate sets and compare gate-level noise, where Pauli channels are inserted after elementary gates, with layer-level noise, where an effective Pauli channel is applied after each \ac{QAOA} layer.
 	The differently compiled circuits are not used to investigate which compilation is better, but to make sure that the observed asymmetry is not a compilation-specific effect.
 	The noisy evolution is modeled using standard Pauli channels that have been used to study local noise in \ac{QAOA} circuits \cite{Marshall_2020}. 
 	The layer level model treats the accumulated noise of one \ac{QAOA} layer as an effective channel, while the gate level noise resolves how errors accumulate after the individual operations required by a particular compilation. 
 	Finally we use the observed asymmetry between bit-flip and phase-flip errors to motivate the use of asymmetric quantum error correcting codes tailored to the dominant error channel.  
 	
 	In \cref{sec:background} we introduce the example application that is used in our analysis.
 	Three different levels of compilation and the noise model are introduced in \cref{sec:noise_modeling} and used in our simulation in \cref{sec:noise_simulation}.
 	We discuss the effect of asymmetric quantum error correction in \cref{sec:error_correction}. Finally we conclude the presentation of our study in \cref{sec:conclusion}.
 	
 	\section{Background}
 	\label{sec:background}
 	Many combinatorial optimization problems are computationally hard, although important special cases admit polynomial-time algorithms, such as the shortest path problem on a weighted graph \cite{Dijkstra}. 
 	Indeed, many scheduling, assignment, routing, and constraint-satisfaction problems are \ac{NP}-hard, including MaxCut, 3-SAT, and variants of the \ac{FGA} problem \cite{ausiello2012complexity}. 
 	Such problems are characterized by a discrete solution space, where the goal is to find an assignment of decision variables that satisfies the problem constraints and optimizes a given cost function.
 	A general constrained combinatorial optimization problem can be written as
 	\begin{align}
 		\min_{x \in \mathcal{F}} C(x),
 		\label{eq:combinatorial}
 	\end{align}
 	where $x=(x_1,x_2,\ldots,x_n)$ denotes a vector of discrete decision variables and $\mathcal{F}$ is the set of feasible assignments. 
 	In many applications, the variables are binary, $x_i\in\{0,1\}$, such that $\mathcal{F}\subseteq\{0,1\}^n$. 
 	The feasible set is determined by the constraints of the problem. 
 	For example, in a graph-coloring problem adjacent vertices must not be assigned the same color, while in a traveling-salesperson problem each city must be visited exactly once. 
 	In the flight-gate assignment problem considered in this work, each flight must be assigned to exactly one gate, and two temporally conflicting flights cannot be assigned to the same gate. 
 	For most of the cases finding the set $\mathcal{F}$ is a non-trivial task, and  the number of feasible or candidate assignments can grow exponentially with the number of decision variables. 
 	A common strategy is to reformulate the constrained problem as a \ac{QUBO} by adding penalty terms to the cost function.
 	This leads naturally to a \ac{QUBO} formulation, which can then be mapped to an Ising Hamiltonian for use in \ac{QAOA} \cite{Lucas:2014}.

 	\subsection{Instance Definition}
 	According to the \ac{FGA} problem, the flights have to be assigned to available airport gates such that passenger walking and transfer times are minimized. The total passenger time depends on three main contributions: the time required by departing passengers to reach their assigned gate, time required by arriving passengers to reach to the baggage claim, and the transfer time of the passengers changing from one flight to another. In addition, each assignment must satisfy the constraints that each flight has to be assigned to exactly one gate and no two flights whose gate-occupation time overlap cannot be assigned to the same gate.
 	
 	We define the problem using the following sets and parameters shown in \cref{tab:instance-definitions}. The binary decision variable  $x_{i \alpha}$ is defined as 
 	\begin{align}
 		x_{i \alpha} ={}&
 		\begin{cases}
 			1, & \text{if flight} \ i \ \text{is assigned to gate } \alpha, \\
 			0, & \text{otherwise.} 
 		\end{cases}
 	\end{align}
 	Thus an assignment is represented by $x \in \{0, 1\}^{F \times G}$. 
 	
 	\begin{table*}[tbph]
 		\caption{Data Instance Definitions for the \ac{FGA} Problem. }
 		\centering
 		\begin{tabular}{|p{1.3cm}|p{6.0cm}|p{4.5cm}|}
 			\hline
 			\textbf{Symbol} & \textbf{Description} & \textbf{Simulated Instance}\\
 			\hline
 			$F$ & Set of flights; $i, j$ refer to elements in $F$ & $\{F_1, F_2, F_3\}$\\
 			\hline
 			$G$ & Set of gates; $\alpha, \beta$ refer to elements in $G$ & $\{G_1, G_2\}$ \\
 			\hline
 			$n_i^d$ & Number of passengers departing on flight $i$ & $n^d_1 =37, n^d_2 =145, n^d_3 =45 $ \\
 			\hline
 			$t^d_\alpha$ & Time from check-in to gate $\alpha$ & $t^d_1 =1270.16 s, t^d_2 =1263.83 s$\\
 			\hline
 			$n_i^a$ & Number of passengers arriving with flight $i$ &  $n^a_1 =33, n^a_2 =152, n^a_3 =0 $ \\
 			\hline
 			$t^a_\alpha$ & Time from gate $\alpha$ to baggage claim & $t^a_1 =1270.16 s, t^a_2 =1263.83 s$ \\
 			\hline
 			$n_{ij}$ & Number of layover passengers from flight $i$ to $j$ & $n_{1,2} = 7, n_{1, 3} =6, n_{2,3} =0$ \\
 			\hline
 			$t_{\alpha\beta}$ & Time it takes to move from gate $\alpha$ to gate $\beta$ & $t_{1,2} = t_{2,1} = 300s$\\
 			\hline
 			$t_i^{\text{in}}$ & Arrival time of flight $i$ &  ${t^{\text{in}}_1=06:00:00}$, ${t^{\text{in}}_2=09:05:00}$, ${t^{\text{in}}_3=11:04:58}$\\
 			\hline
 			$t_i^{\text{out}}$ & Departure time of flight $i$ &  ${t^{\text{out}}_1 = 07:39:59}$, ${t^{\text{out}}_2 =12:44:57}$, ${t^{\text{out}}_3 = 13:04:58}$ \\
 			\hline
 			$t^{\text{buf}}$ & Buffer time between two flights at the same gate & $t^{\text{buf}} = 500 s$ \\
 			\hline
 		\end{tabular}
 		\label{tab:instance-definitions}
 	\end{table*}
 	
 	\subsubsection{Objective Function}
 	For a given assignment $x$, the total passenger walking and transfer time is given by  
 	\begin{align}
 		C(x) = \sum_{i,\alpha} (n_i^d t^d_\alpha + n_i^a t^a_\alpha) x_{i\alpha} 
 		+ \sum_{i,j,\alpha,\beta} n_{ij} t_{\alpha\beta} x_{i\alpha} x_{j\beta}.
 		\label{eq:objective}
 	\end{align} 
 	The fist term accounts for passengers departing from and arriving at the airport. For each flight assigned to gate $\alpha$, it adds the time required by departing passengers to move from check-in to that gate and by arriving passengers to move from that gate to baggage claim. The second term accounts for transfer passengers changing from flight $i$ to flight $j$. It is quadratic in assignment variable because it depends on pair of gates assigned to two flights.
 	
 	The function $C(x)$ only represent the passengers commuting time objective function, and does not by itself enforce the feasibility of the assignment. For example, minimizing $C(x)$ without constraints could assign several flights to the same gate or assign a flight to more than one gate. The feasible assignments are therefore restricted by two constraints. First, each flight must be assigned to exactly one gate, 
 	\begin{equation}
 		\sum_{\alpha \in G} x_{i\alpha} = 1, \quad \forall i \in F.
 		\label{eq:unique_gate}
 	\end{equation}
 	Second, two conflicting flights cannot be assigned to the same gate, 
 	\begin{equation}
 		x_{i\alpha} x_{j\alpha} = 0, \quad \forall (i, j) \in P, \forall \alpha \in G.
 		\label{eq:time_conflict}
 	\end{equation}
 	Here, $P$ denotes the set of flight pairs with overlapping gate-occupation intervals and is defined as
 	\begin{equation}
 		P = \left\{ (i,j) \in F^2 \;\middle|\; t_i^{\text{in}} < t_j^{\text{out}} < t_i^{\text{out}} + t^{\text{buf}} \right\}.
 		\label{eq:conflict_set}
 	\end{equation}
 	Thus, if $(i,j)\in P$, the two flights should not be assigned to the same gate.

 	\subsection{\ac{QUBO} Formulation}
 	A \ac{QUBO} problem is an optimization problem over binary variables whose objective function contains linear and quadratic terms. For a binary vector $x \in \{ 0, 1\}^n$, the general \ac{QUBO} objective can be written as 
 	\begin{align}
 		Q(x) = x^{T}Qx = \sum_{i}^n Q_{ii}x_i+ \sum_{\stackrel{i,j=1}{i < j}}^nQ_{ij}x_ix_j. 
 	\end{align}
 	The diagonal coefficients $Q_{ii}$  define the linear contribution of each binary variable, while the off-diagonal coefficients $Q_{ij}$ define pairwise interactions between variables. The constrained \ac{FGA} problem can be converted into \ac{QUBO} form by adding penalty terms to the passenger--time objective. Instead of minimizing $C(x)$ only over the feasible set $\mathcal{F}$, we minimize an un constrained objective over all binary assignments, 
 	\begin{align}
 		\min_{x\in\{0,1\}^{F\times G}} Q(x).
 		\label{eq:qubo_optimization}
 	\end{align}
 	The constraints are incorporated into the objective function by assigning a positive penalty to invalid assignments. 
 	For the \ac{FGA} problem, we use the \ac{QUBO} objective
 	\begin{align}
 		Q(x) = C(x) + \lambda_u C_u(x) + \lambda_t C_t(x),
 		\label{eq:qubo}
 	\end{align}
 	where $C_u(x)$ penalizes violations of the unique-gate constraint and $C_t(x)$ penalizes assignments in which conflicting flights use the same gate. The unique-gate penalty is given by
 	\begin{align}
 		C_u(x)
 		=
 		\sum_{i\in F}
 		\left(
 		\sum_{\alpha\in G} x_{i\alpha} - 1
 		\right)^2 ,
 		\label{eq:unique_penalty}
 	\end{align}
 	which vanishes only when each flight is assigned to exactly one gate. 
 	The conflict penalty is given by
 	\begin{align}
 		C_t(x)
 		=
 		\sum_{\alpha\in G}
 		\sum_{(i,j)\in P}
 		x_{i\alpha}x_{j\alpha},
 		\label{eq:conflict_penalty}
 	\end{align}
 	which increases the objective whenever two conflicting flights are assigned to the same gate. 
 	The positive coefficients $\lambda_u$ and $\lambda_t$ control the relative strength of the two penalty terms. 
 	For sufficiently large penalty weights, the cost function for infeasible assignments become unfavorably large compared with feasible assignments of the original problem.
 	
 	\subsubsection{Penalty Weight Estimation}
 	The choice of penalty weights is important because the penalties must enforce feasibility without unnecessarily distorting the optimization landscape. 
 	If the penalty weights are too small, the optimizer may prefer infeasible assignments with low passenger time. 
 	If they are too large, the penalty terms can dominate the objective and make the optimization landscape harder to explore. 
 	In this work, we use penalty weights previously used in \ac{DLR} studies of the \ac{FGA} problem with classical optimizers \cite{quark2024}.
 	
 	The penalty coefficient for the unique-gate constraint is chosen as
 	\begin{align}
 		\lambda_u^{\mathrm{opt}}
 		=
 		\max_{i,\alpha}
 		\left[
 		n_i^d t_\alpha^d
 		+
 		n_i^a t_\alpha^a
 		+
 		\left(
 		\max_{\alpha,\beta} t_{\alpha\beta}
 		\right)
 		\sum_j n_{ij}
 		\right],
 		\label{eq:lambda_u}
 	\end{align}
 	and the penalty coefficient for the conflict constraint is chosen as
 	\begin{align}
 		\lambda_t^{\mathrm{opt}}
 		={}&
 		\max_{i,\alpha,\gamma}
 		\Bigg[
 		\left(n_i^d t_\alpha^d + n_i^a t_\alpha^a\right)
 		-
 		\left(n_i^d t_\gamma^d + n_i^a t_\gamma^a\right) \nonumber \\
 		&+ \left(
 		\max_{\alpha,\beta} t_{\alpha\beta}
 		- \min_{\alpha,\beta} t_{\alpha\beta}
 		\right)
 		\sum_j n_{ij}
 		\Bigg].
 		\label{eq:lambda_t}
 	\end{align}

 	For the simulated instance in \cref{tab:instance-definitions}, the penalty-weight
 	estimates obtained from Eq.~\eqref{eq:lambda_u} and Eq.~\eqref{eq:lambda_t} are
 	$\lambda_u^{\mathrm{opt}}=377239.5$ and
 	$\lambda_t^{\mathrm{opt}}=1881.0$, respectively.
 	To provide an additional margin for enforcing the
 	assignment constraints, we choose slightly larger
 	penalty weights,
 	\begin{align}
 		\lambda_u=377439.5,
 		\qquad
 		\lambda_t=2081.0.
 		\label{eq:penalty_weights}
 	\end{align}
 	This choice penalizes constraint violations more strongly relative to the passenger-time objective, thereby favoring feasible assignments. Both penalty weights are given in the same passenger-time units as the objective function and are used to construct the penalized QUBO in Eq.~\eqref{eq:qubo}.

 	\subsection{Ising Hamiltonians of the \ac{QAOA}}
 	\label{subsec:ising_mapping}

 	To implement the \ac{QUBO} objective in \ac{QAOA}, the binary variables are mapped to spin variables and then to Pauli operators. For a binary variable $x_k\in\{0,1\}$, we use the standard transformation \begin{align} x_k = \frac{1-s_k}{2}, \qquad s_k \in \{-1,+1\}. \end{align} At the operator level, the spin variable $s_k$ is represented by the Pauli--$Z$ operator acting on qubit $k$, so that \begin{align} x_k \longmapsto \frac{I-Z_k}{2}. \end{align} With this mapping, the \ac{QUBO} objective in Eq.~\eqref{eq:qubo} becomes a diagonal Ising Hamiltonian of the form 
 	\begin{align}
 		H_{\mathrm{cost}} = \alpha_0 I + \sum_k h_k Z_k + \sum_{k<l} J_{kl} Z_k Z_l . \label{eq:general_ising} 
 	\end{align}
 	The constant $\alpha_0$ shifts all energies by the same amount and therefore does not affect the location of the optimal bit string. The coefficients $h_k$ and $J_{kl}$ are determined by the linear and quadratic \ac{QUBO} coefficients after applying the binary-to-spin transformation. 
 	In the present problem, the binary index $k$ corresponds to a flight--gate assignment variable $x_{i\alpha}$. Therefore, an instance with $|F|$ flights and $|G|$ gates requires $|F||G|$ binary variables and hence the same number of qubits. The ground state of the final cost Hamiltonian $H_C$ encodes the assignment that minimizes the penalized objective function, provided that the penalty weights are chosen large enough to suppress infeasible assignments. 
 	For the numerical study in this work, we use a reduced \ac{FGA} instance derived from cleaned airport data considered in Ref.~\cite{Elisabeth_FGA}. 
 	The instance contains three flights and two gates, resulting in six binary assignment variables and therefore six qubits. 
 	Applying the mapping above gives the following Hamiltonian contributions.
 	
 	\paragraph{Passenger-time Hamiltonian.}  
 	The Hamiltonian corresponding to the passenger walking and transfer-time objective is
 	\begin{align}
 		H_{\mathrm{cost}}
 		={}&525904\,I \nonumber\\
 		&-29478.75\,Z_0 
 		-29336.25\,Z_1 
 		-189669.75\,Z_2 \nonumber\\
 		& -188729.25\,Z_3 
 		-46405.8333\,Z_4 
 		-46184.1667\,Z_5 \nonumber\\
 		&+450\left(
 		Z_0Z_4+Z_0Z_5+Z_1Z_4+Z_1Z_5
 		\right)
 		\nonumber\\
 		&+525\left(
 		Z_2Z_4+Z_2Z_5+Z_3Z_4+Z_3Z_5
 		\right).
 		\label{eq:passenger_time_hamiltonian}
 	\end{align}

 	This Hamiltonian represents the objective contribution before adding penalties for invalid assignments.
 	
 	\paragraph{Unique-assignment constraint Hamiltonian.} 
 	The penalty enforcing that each flight is assigned to exactly one gate is mapped to
 	\begin{align}
 		H_u=
 		1.5I
 		+
 		0.5
 		\left(
 		Z_0Z_1+Z_2Z_3+Z_4Z_5
 		\right).
 		\label{eq:unique_assignment_hamiltonian}
 	\end{align}
 	
 	Each two-qubit term acts on the pair of gate-assignment variables corresponding to the same flight. 
 	Thus, this Hamiltonian penalizes configurations in which a flight is either assigned to no gate or to more than one gate.
 	
 	\paragraph{Conflict-avoidance constraint Hamiltonian.} 
 	The penalty enforcing that two conflicting flights are not assigned to the same gate is mapped to
 	\begin{align}
 		H_t
 		={}&
 		0.5I
 		-0.25
 		\left(
 		Z_0+Z_1+Z_2+Z_3 -Z_0Z_2-Z_1Z_3
 		\right).
 		\label{eq:conflict_hamiltonian}
 	\end{align}
 	
 	The two-qubit terms couple assignment variables that correspond to conflicting flights being placed at the same gate. 
 	
 	Combining these contributions with the penalty
 	weights from Eq.~\eqref{eq:penalty_weights} gives the
 	unnormalized cost Hamiltonian,
 	\begin{align}
 		\widetilde H_C
 		=
 		H_{\mathrm{cost}}
 		+\lambda_u H_u
 		+\lambda_t H_t.
 		\label{eq:combined_hamiltonian}
 	\end{align}
 	For the numerical simulations, the complete Hamiltonian
 	is normalized by its largest absolute Pauli coefficient,
 	\begin{align}
 		H_C=\frac{\widetilde H_C}{s_C},
 		\label{eq:normalized_hamiltonian}
 	\end{align}
 	where $s_C$ denotes the largest absolute coefficient
 	in the Pauli expansion of $\widetilde H_C$.
 	This normalization preserves the relative weights
 	of the objective and constraint terms and does not
 	change the minimizing computational-basis states.


 	\paragraph{Mixer Hamiltonian} We use the standard transverse-field mixer \begin{align} H_M = \sum_{j=0}^{5} X_j , \label{eq:mixer_hamiltonian} \end{align} where $X_j$ is the Pauli--$X$ operator acting on qubit $j$. The mixer does not commute with the cost Hamiltonian and therefore allows the \ac{QAOA} circuit to create transitions between different assignment bit strings.
 	
 	\section{\ac{QAOA} Circuit Decomposition and Noise Models}
 	\label{sec:noise_modeling}
 	
 	The \ac{QAOA} ansatz at the level of  Hamiltonian evolutions is generated by alternating the cost Hamiltonian and the mixer Hamiltonian. 
 	For a noiseless circuit, these evolutions define a unitary map from the initial state to the final variational state. 
 	For a noisy circuit simulation, however, the same logical unitary has to be implemented using elementary gates, and noise channels must be inserted at a chosen level of the circuit. 
 	In this section, we first define the ideal \ac{QAOA} evolution, then discuss its decomposition into two different gate sets, and finally introduce the Pauli noise channels used in the gate-level and layer-level simulations.
 	
 	\subsection{Ideal \ac{QAOA} Ansatz as a Unitary Evolution}
 	\label{subsec:ideal_qaoa_unitary}
 	
 	We first review the ideal \ac{QAOA} circuit without noise. 
 	A depth-$d$ \ac{QAOA} ansatz is specified by two real parameter vectors,
 	\begin{align}
 		\boldsymbol{\gamma}
 		=
 		(\gamma_1,\gamma_2,\ldots,\gamma_d),
 		\qquad
 		\boldsymbol{\beta}
 		=
 		(\beta_1,\beta_2,\ldots,\beta_d).
 	\end{align}
 	The parameter $\gamma_\ell$ controls the evolution under the cost Hamiltonian in layer $\ell$, while $\beta_\ell$ controls the evolution under the mixer Hamiltonian in the same layer.
 	
 	For the $\ell$-th \ac{QAOA} layer, the corresponding cost and mixer unitaries are defined as
 	\begin{align}
 		U_C(\gamma_\ell)
 		&=
 		\exp\!\left(-i\gamma_\ell H_C\right),\\
 		\qquad
 		U_M(\beta_\ell)
 		&=
 		\exp\!\left(-i\beta_\ell H_M\right),
 		\label{eq:qaoa_layer_unitaries}
 	\end{align}
 	where $H_C$ is the full cost Hamiltonian used for the optimization and $H_M$ is the mixer Hamiltonian. 
 	In our case, $H_C$ is diagonal in the computational basis, while $H_M$ is generated by Pauli--$X$ operators. 
 	Since these two Hamiltonians do not commute, their alternating application creates a nontrivial variational family of quantum states.
 	
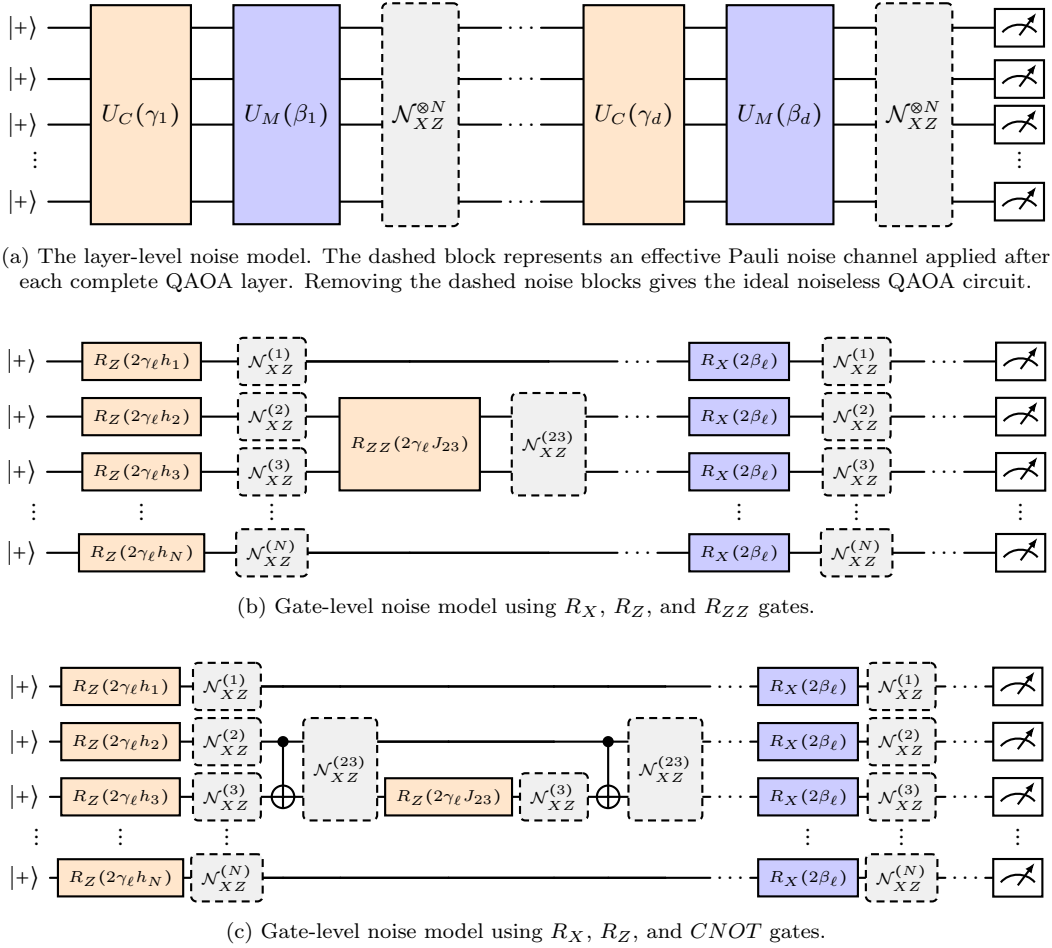
\begin{figure*}[ptbh]
 		\centering
 		
 		
 		\subfloat[The layer-level noise model. The dashed block represents an effective Pauli noise channel applied after each complete QAOA layer. Removing the dashed noise blocks gives the ideal noiseless QAOA circuit.\label{fig:circuits_layers}]{%
 			\begin{quantikz}[wire types={q,q,q,n,q}, column sep=13.5pt, row sep=3pt]
 				\lstick{$\ket{+}$} & \gate[5,style={fill=orange!20}]{U_C(\gamma_1)} & \gate[5,style={fill=blue!20}]{U_M(\beta_1)} & \gate[5,style={draw=black,densely dashed,fill=gray!12,rounded corners=2pt}]{\mathcal{N}^{\otimes N}_{XZ}} & \push{\;\dots\;} & \gate[5,style={fill=orange!20}]{U_C(\gamma_d)} & \gate[5,style={fill=blue!20}]{U_M(\beta_d)} & \gate[5,style={draw=black,densely dashed,fill=gray!12,rounded corners=2pt}]{\mathcal{N}^{\otimes N}_{XZ}} & \meter{} \\
 				\lstick{$\ket{+}$} & & & & \push{\;\dots\;} & & & & \meter{} \\
 				\lstick{$\ket{+}$} & & & & \push{\;\dots\;} & & & & \meter{} \\
 				\lstick{$\vdots$} & & & &  & & & & \vdots \\
 				\lstick{$\ket{+}$} & & & & \push{\;\dots\;} & & & & \meter{}
 			\end{quantikz}%
 		}
 		
 		
 		\subfloat[Gate-level noise model using $R_X$, $R_Z$, and $R_{ZZ}$ gates.\label{fig:circuits_native}]{%
 			\begin{quantikz}[wire types={q,q,q,n,q}, column sep=12pt, row sep=3pt]
 				\lstick{$\ket{+}$} & \gate[style={fill=orange!20}]{\scriptstyle R_Z(2\gamma_\ell h_1)} & \gate[style={draw=black,densely dashed,fill=gray!12,rounded corners=2pt}]{\scriptstyle \mathcal{N}^{(1)}_{XZ}} & \qw & \qw & \push{\;\dots\;} & \gate[style={fill=blue!20}]{\scriptstyle R_X(2\beta_\ell)} & \gate[style={draw=black,densely dashed,fill=gray!12,rounded corners=2pt}]{\scriptstyle \mathcal{N}^{(1)}_{XZ}} & \push{\;\dots\;} & \meter{} \\
 				\lstick{$\ket{+}$} & \gate[style={fill=orange!20}]{\scriptstyle R_Z(2\gamma_\ell h_2)} & \gate[style={draw=black,densely dashed,fill=gray!12,rounded corners=2pt}]{\scriptstyle \mathcal{N}^{(2)}_{XZ}} & \gate[2,style={fill=orange!20}]{\scriptstyle R_{ZZ}(2\gamma_\ell J_{23})} & \gate[2,style={draw=black,densely dashed,fill=gray!12,rounded corners=2pt}]{\scriptstyle \mathcal{N}^{(23)}_{XZ}} & \push{\;\dots\;} & \gate[style={fill=blue!20}]{\scriptstyle R_X(2\beta_\ell)} & \gate[style={draw=black,densely dashed,fill=gray!12,rounded corners=2pt}]{\scriptstyle \mathcal{N}^{(2)}_{XZ}} & \push{\;\dots\;} & \meter{} \\
 				\lstick{$\ket{+}$} & \gate[style={fill=orange!20}]{\scriptstyle R_Z(2\gamma_\ell h_3)} & \gate[style={draw=black,densely dashed,fill=gray!12,rounded corners=2pt}]{\scriptstyle \mathcal{N}^{(3)}_{XZ}} & & & \push{\;\dots\;} & \gate[style={fill=blue!20}]{\scriptstyle R_X(2\beta_\ell)} & \gate[style={draw=black,densely dashed,fill=gray!12,rounded corners=2pt}]{\scriptstyle \mathcal{N}^{(3)}_{XZ}} & \push{\;\dots\;} & \meter{} \\
 				\lstick{$\vdots$} & \vdots & \vdots & & &  & \vdots & \vdots &  & \vdots \\
 				\lstick{$\ket{+}$} & \gate[style={fill=orange!20}]{\scriptstyle R_Z(2\gamma_\ell h_N)} & \gate[style={draw=black,densely dashed,fill=gray!12,rounded corners=2pt}]{\scriptstyle \mathcal{N}^{(N)}_{XZ}} & \qw & \qw & \push{\;\dots\;} & \gate[style={fill=blue!20}]{\scriptstyle R_X(2\beta_\ell)} & \gate[style={draw=black,densely dashed,fill=gray!12,rounded corners=2pt}]{\scriptstyle \mathcal{N}^{(N)}_{XZ}} & \push{\;\dots\;} & \meter{}
 			\end{quantikz}%
 		}\\
 		
 		\subfloat[Gate-level noise model using $R_X$, $R_Z$, and $CNOT$ gates.\label{fig:circuits_decomposed}]{%
 			\begin{quantikz}[wire types={q,q,q,n,q}, column sep=3pt, row sep=3pt]
 				\lstick{$\ket{+}$} & \gate[style={fill=orange!20}]{\scriptstyle R_Z(2\gamma_\ell h_1)} & \gate[style={draw=black,densely dashed,fill=gray!12,rounded corners=2pt}]{\scriptstyle \mathcal{N}^{(1)}_{XZ}} & \qw & \qw & \qw & \qw & \qw & \qw & \push{\;\dots\;} & \gate[style={fill=blue!20}]{\scriptstyle R_X(2\beta_\ell)} & \gate[style={draw=black,densely dashed,fill=gray!12,rounded corners=2pt}]{\scriptstyle \mathcal{N}^{(1)}_{XZ}} & \push{\;\dots\;} & \meter{} \\
 				\lstick{$\ket{+}$} & \gate[style={fill=orange!20}]{\scriptstyle R_Z(2\gamma_\ell h_2)} & \gate[style={draw=black,densely dashed,fill=gray!12,rounded corners=2pt}]{\scriptstyle \mathcal{N}^{(2)}_{XZ}} & \ctrl{1} & \gate[2,style={draw=black,densely dashed,fill=gray!12,rounded corners=2pt}]{\scriptstyle \mathcal{N}^{(23)}_{XZ}} & \qw & \qw & \ctrl{1} & \gate[2,style={draw=black,densely dashed,fill=gray!12,rounded corners=2pt}]{\scriptstyle \mathcal{N}^{(23)}_{XZ}} & \push{\;\dots\;} & \gate[style={fill=blue!20}]{\scriptstyle R_X(2\beta_\ell)} & \gate[style={draw=black,densely dashed,fill=gray!12,rounded corners=2pt}]{\scriptstyle \mathcal{N}^{(2)}_{XZ}} & \push{\;\dots\;} & \meter{} \\
 				\lstick{$\ket{+}$} & \gate[style={fill=orange!20}]{\scriptstyle R_Z(2\gamma_\ell h_3)} & \gate[style={draw=black,densely dashed,fill=gray!12,rounded corners=2pt}]{\scriptstyle \mathcal{N}^{(3)}_{XZ}} & \targ{} & & \gate[style={fill=orange!20}]{\scriptstyle R_Z(2\gamma_\ell J_{23})} & \gate[style={draw=black,densely dashed,fill=gray!12,rounded corners=2pt}]{\scriptstyle \mathcal{N}^{(3)}_{XZ}} & \targ{} & & \push{\;\dots\;} & \gate[style={fill=blue!20}]{\scriptstyle R_X(2\beta_\ell)} & \gate[style={draw=black,densely dashed,fill=gray!12,rounded corners=2pt}]{\scriptstyle \mathcal{N}^{(3)}_{XZ}} & \push{\;\dots\;} & \meter{} \\
 				\lstick{$\vdots$} & \vdots & \vdots & & & & & & &  & \vdots & \vdots &  & \vdots \\
 				\lstick{$\ket{+}$} & \gate[style={fill=orange!20}]{\scriptstyle R_Z(2\gamma_\ell h_N)} & \gate[style={draw=black,densely dashed,fill=gray!12,rounded corners=2pt}]{\scriptstyle \mathcal{N}^{(N)}_{XZ}} & \qw & \qw & \qw & \qw & \qw & \qw & \push{\;\dots\;} & \gate[style={fill=blue!20}]{\scriptstyle R_X(2\beta_\ell)} & \gate[style={draw=black,densely dashed,fill=gray!12,rounded corners=2pt}]{\scriptstyle \mathcal{N}^{(N)}_{XZ}} & \push{\;\dots\;} & \meter{}
 			\end{quantikz}%
 		}%
 		
 		\caption{The \ac{QAOA} circuit with the noise insertions as described in Sec. \ref{subsec:pauli_noise_channels}. Panel~\ref{fig:circuits_layers} shows the layer-level description, where one effective Pauli noise channel $\mathcal{N}^{\otimes N}_{XZ}$ is applied after each complete QAOA layer. Panels~\ref{fig:circuits_native} and~\ref{fig:circuits_decomposed} show representative gate-level implementations of a layer $\ell$, with $1<\ell<d$, for two different gate sets. Here $\mathcal{N}^{(j)}_{XZ}$ denotes the single-qubit bit-flip/phase-flip Pauli channel applied after a single-qubit gate on qubit $j$, while $\mathcal{N}^{(ij)}_{XZ}$ denotes the corresponding two-qubit Pauli channel applied after a two-qubit gate acting on qubits $i$ and $j$. In the native implementation, the $ZZ$ interaction is implemented directly by $R_{ZZ}(2\gamma_\ell J_{23})$, followed by the two-qubit noise channel $\mathcal{N}^{(23)}_{XZ}$. In the decomposed implementation, the same interaction is realized by a $CNOT$--$R_Z$--$CNOT$ sequence, leading to additional noisy locations. Removing all dashed noise boxes recovers the corresponding ideal noiseless QAOA circuits.}
 		\label{fig:circuits}
 	\end{figure*}

 	The full depth-$d$ \ac{QAOA} unitary is obtained by applying the cost and mixer evolutions alternately, as shown in \cref{fig:circuits_layers}. In \cref{fig:circuits}, the solid gates represent the underlying ideal circuit. The ideal noiseless case is recovered by omitting the dashed noise channels, or equivalently by setting the corresponding noise probabilities to zero. With the convention that the rightmost operator acts first, the depth-$d$ unitary is
 	\begin{equation}
 		\begin{aligned}
 			U_d(\boldsymbol{\gamma},\boldsymbol{\beta})
 			={}&
 			U_M(\beta_d)U_C(\gamma_d)
 			\cdots  \\
 			&
 			U_M(\beta_2)U_C(\gamma_2)
 			U_M(\beta_1)U_C(\gamma_1).
 		\end{aligned}
 		\label{eq:qaoa_unitary}
 	\end{equation}
 	Thus, each \ac{QAOA} layer consists of a cost evolution followed by a mixer evolution. 
 	This convention is used throughout the numerical simulations.
 	
 	The standard \ac{QAOA} initial state is the uniform superposition over all computational basis states. 
 	For an $n$-qubit problem, it is written as
 	\begin{equation}
 		\rho_0
 		=
 		\left(\ket{+}\bra{+}\right)^{\otimes n},
 		\qquad
 		\ket{+}
 		=
 		\frac{1}{\sqrt{2}}\left(\ket{0}+\ket{1}\right).
 		\label{eq:qaoa_initial_state}
 	\end{equation}
 	For the six-qubit \ac{FGA} instance considered in this work, $n=6$.
 	
 	After applying the depth-$d$ \ac{QAOA} unitary, the ideal final state is
 	\begin{align}
 		\rho_d
 		=
 		U_d(\boldsymbol{\gamma},\boldsymbol{\beta})
 		\rho_0
 		U_d^\dagger(\boldsymbol{\gamma},\boldsymbol{\beta}).
 		\label{eq:ideal_qaoa_state}
 	\end{align}
 	Equivalently, the noiseless \ac{QAOA} evolution can be written as the unitary quantum channel
 	\begin{align}
 		\mathcal{U}_d(\rho)
 		=
 		U_d(\boldsymbol{\gamma},\boldsymbol{\beta})
 		\rho
 		U_d^\dagger(\boldsymbol{\gamma},\boldsymbol{\beta}).
 		\label{eq:ideal_qaoa_channel}
 	\end{align}
 	
 	The variational objective optimized by \ac{QAOA} is the expectation value of the cost Hamiltonian with respect to the final state,
 	\begin{align}
 		C_d(\boldsymbol{\gamma},\boldsymbol{\beta})
 		=
 		\operatorname{Tr}
 		\left[
 		H_C \rho_d
 		\right]
 		=
 		\operatorname{Tr}
 		\left[
 		H_C\,\mathcal{U}_d(\rho_0)
 		\right].
 		\label{eq:ideal_qaoa_cost}
 	\end{align}
 	The classical optimizer updates the parameters $\boldsymbol{\gamma}$ and $\boldsymbol{\beta}$ in order to minimize this expectation value. 
 	This ideal unitary evolution serves as the reference case against which the noisy gate-level and layer-level simulations are compared.

 	\subsection{Gate-Set Decompositions of the \ac{QAOA} Layer}
 	\label{subsec:gate_sets}
 	
 	The ideal \ac{QAOA} layer in Eq.~\eqref{eq:qaoa_layer_unitaries} is defined at the Hamiltonian level. 
 	For a circuit-level implementation, the cost and mixer unitaries have to be decomposed into elementary gates. 
 	This decomposition is not unique. 
 	Different gate sets can implement the same ideal unitary, but they may require different numbers of single-qubit and two-qubit gates. 
 	This distinction is important for the gate-level noise model, because noise channels are inserted after the elementary gates of the compiled circuit.
 	
 	We use the standard gate conventions
 	\begin{align}
 		R_z(\theta)
 		={}&
 		\exp\!\left(-i\frac{\theta}{2}Z\right),
 		\qquad
 		R_x(\theta)
 		=
 		\exp\!\left(-i\frac{\theta}{2}X\right),
 		\qquad  \nonumber \\
 		&\text{and } R_{ZZ}(\theta)
 		=
 		\exp\!\left(-i\frac{\theta}{2}Z\otimes Z\right).
 		\label{eq:rotation_conventions}
 	\end{align}
 	With these definitions, the factors of two in the rotation angles arise when the Hamiltonian evolutions are written as elementary rotation gates.
 	
 	The cost Hamiltonian is of Ising form
 	\begin{align}
 		H_C
 		=
 		\alpha_0 I
 		+
 		\sum_k h_k Z_k
 		+
 		\sum_{k<l}J_{kl}Z_kZ_l .
 		\label{eq:ising_cost_for_decomposition}
 	\end{align}
 	Since all $Z_k$ and $Z_kZ_l$ terms commute, the cost unitary in layer $\ell$ factorizes as
 	\begin{align}
 		U_C(\gamma_\ell)
 		={}&
 		\exp(-i\gamma_\ell \alpha_0)
 		\prod_k
 		\exp(-i\gamma_\ell h_k Z_k) \nonumber \\
 		&\prod_{k<l}
 		\exp(-i\gamma_\ell J_{kl}Z_kZ_l).
 		\label{eq:cost_unitary_factorized}
 	\end{align}
 	The global phase $\exp(-i\gamma_\ell\alpha_0)$ has no effect on measurement probabilities or expectation values and is therefore omitted in the circuit implementation. 
 	The remaining single-qubit and two-qubit terms are implemented as
 	\begin{align}
 		\exp(-i\gamma_\ell h_k Z_k)
 		=
 		R_z(2\gamma_\ell h_k),
 		\label{eq:z_rotation_mapping}
 	\end{align}
 	and
 	\begin{align}
 		\exp(-i\gamma_\ell J_{kl}Z_kZ_l)
 		=
 		R_{ZZ}^{(kl)}(2\gamma_\ell J_{kl}).
 		\label{eq:zz_rotation_mapping}
 	\end{align}
 	
 	The mixer Hamiltonian is
 	\begin{align}
 		H_M
 		=
 		\sum_k X_k .
 	\end{align}
 	Since the Pauli--$X$ operators on different qubits commute, the mixer unitary also factorizes into single-qubit rotations,
 	\begin{align}
 		U_M(\beta_\ell)
 		=
 		\prod_k
 		\exp(-i\beta_\ell X_k)
 		=
 		\prod_k
 		R_x(2\beta_\ell).
 		\label{eq:mixer_unitary_factorized}
 	\end{align}
 	Thus, the mixer part is implemented in the same way in both gate sets considered here.
 	
 	In this work, we compare two circuit decompositions of the same logical \ac{QAOA} layer. 
 	The first gate set is
 	\begin{align}
 		\mathcal{G}_1
 		=
 		\{R_z, R_x, \mathrm{CNOT}\}.
 		\label{eq:gateset_1}
 	\end{align}
 	In this decomposition, the single-qubit $Z$ and $X$ rotations are implemented directly as $R_z$ and $R_x$ gates, 
 	while each $ZZ$ rotation is decomposed into a CNOT--$R_z$--CNOT sequence, see also \cref{fig:circuits_decomposed}. 
 	Using qubit $k$ as the control and qubit $l$ as the target, this decomposition is
 	\begin{align}
 		R_{ZZ}^{(kl)}(\theta)
 		=
 		\mathrm{CNOT}_{k\rightarrow l}\,
 		R_z^{(l)}(\theta)\,
 		\mathrm{CNOT}_{k\rightarrow l}.
 		\label{eq:rzz_cnot_decomposition}
 	\end{align}
 	Here the rightmost CNOT acts first. 
 	With the angle convention in Eq.~\eqref{eq:rotation_conventions}, this implements
 	\begin{align}
 		R_{ZZ}^{(kl)}(2\gamma_\ell J_{kl})
 		=
 		\exp(-i\gamma_\ell J_{kl}Z_kZ_l).
 	\end{align}
 	
 	The second gate set is
 	\begin{align}
 		\mathcal{G}_2
 		=
 		\{R_z, R_x, R_{ZZ}\}.
 		\label{eq:gateset_2}
 	\end{align}
 	In this decomposition, the two-qubit Ising interaction is treated as a native $R_{ZZ}$ gate. 
 	Therefore, each interaction term $\exp(-i\gamma_\ell J_{kl}Z_kZ_l)$ is implemented directly as $R_{ZZ}^{(kl)}(2\gamma_\ell J_{kl})$, without decomposing it into CNOT gates and an intermediate $R_z$ rotation, see also \cref{fig:circuits_native}.
 	
 	Both gate sets implement the same ideal \ac{QAOA} unitary. 
 	However, they lead to different compiled circuits. 
 	In particular, each nonzero $ZZ$ interaction requires two CNOT gates and one $R_z$ gate in $\mathcal{G}_1$, but only one native $R_{ZZ}$ gate in $\mathcal{G}_2$. 
 	As a result, the number of locations at which gate-level noise can act depends on the chosen gate set. 
 	For this reason, comparing both decompositions allows us to distinguish effects caused by the physical noise channel from effects caused by the circuit compilation.
 	
 	In the ideal noiseless case, the ordering of the commuting $Z$ and $ZZ$ factors in the cost unitary does not affect the final unitary. 
 	For the noisy gate-level simulations, however, the insertion of noise after elementary gates makes the compiled sequence relevant. 
 	Therefore, we use a fixed ordering of the nonzero Ising terms when constructing the circuits for both gate sets.
 	
 	\subsection{Pauli Noise Channels}
 	\label{subsec:pauli_noise_channels}
 	
 	After fixing the ideal circuit decomposition, we introduce the noise channels used in the simulations. 
 	We restrict our analysis to Pauli bit-flip and phase-flip noise, because these two channels allow us to study biased errors in a controlled way. 
 	The bit-flip channel is generated by Pauli--$X$ errors, while the phase-flip channel is generated by Pauli--$Z$ errors. 
 	By varying their error probabilities independently, we can study both symmetric and asymmetric Pauli noise.
 	
 	For any density matrix $\rho$, the single-qubit bit-flip channel with error probability $p_X$ is defined as
 	\begin{align}
 		\mathcal{N}_{X}^{(j)}(\rho)
 		=
 		(1-p_X)\rho
 		+
 		p_X X_j \rho X_j ,
 		\label{eq:single_qubit_bit_flip}
 	\end{align}
 	where $X_j$ denotes the Pauli--$X$ operator acting on qubit $j$. Analogously, the single-qubit phase-flip channel with error probability $p_Z$ is
 	\begin{align}
 		\mathcal{N}_{Z}^{(j)}(\rho)
 		=
 		(1-p_Z)\rho
 		+
 		p_Z Z_j \rho Z_j ,
 		\label{eq:single_qubit_phase_flip}
 	\end{align}
 	where $Z_j$ denotes the Pauli--$Z$ operator acting on qubit $j$. 
 	Both channels are completely positive and trace preserving, and both reduce to the identity channel when the corresponding error probability is zero.
 	
 	For the asymmetric noise simulations, the bit-flip and phase-flip probabilities are varied independently.
 	In the simulations, the two channels are applied successively as
 	\begin{align}
 		\mathcal{N}_{XZ}^{(j)}
 		=
 		\mathcal{N}_{Z}^{(j)}(p_Z)
 		\circ
 		\mathcal{N}_{X}^{(j)}(p_X).
 		\label{eq:single_qubit_asymmetric_pauli}
 	\end{align}
 	Equivalently, this composed channel can be written as
 	\begin{align}
 		\mathcal{N}_{XZ}^{(j)}(\rho)
 		={}&
 		(1-p_X)(1-p_Z)\rho
 		+
 		p_X(1-p_Z)X_j\rho X_j
 		\nonumber\\
 		&+
 		(1-p_X)p_Z Z_j\rho Z_j
 		+
 		p_Xp_Z Y_j\rho Y_j .
 	\end{align}
 	This keeps the bit-flip and phase-flip probabilities independent and includes mixed \(X\)- and \(Z\)-type events at order \(p_Xp_Z\). In the simulations, this allows us to compare how \(X\)-type and \(Z\)-type errors affect the same \ac{QAOA} circuit.
 	
 	For two-qubit gates, we use the corresponding two-qubit Pauli channels acting on the pair of qubits involved in the gate. 
 	For a gate acting on qubits $i$ and $j$, the two-qubit bit-flip channel is
 	\begin{align}
 		\mathcal{N}_{X}^{(ij)}(\rho)
 		={}&
 		(1-p_X)\rho
 		\nonumber \\
 		&+ 
 		\frac{p_X}{3}
 		\left(
 		X_i\rho X_i
 		+
 		X_j\rho X_j
 		+
 		X_iX_j\rho X_iX_j
 		\right).
 		\label{eq:two_qubit_bit_flip}
 	\end{align}
 	This channel distributes the total bit-flip error probability over three possibilities: an $X$ error on the first qubit, an $X$ error on the second qubit, or an $X$ error on both qubits with equal probability. 
 	While this is the same idea as for the single qubit noise channel, note that this choice implies that the marginal of the two-qubit channel is not the single qubit noise channel and in this sense $p_X$ has a different meaning for single and two-qubit gates.
 	Analogously, the two-qubit phase-flip channel is
 	\begin{align}
 		\mathcal{N}_{Z}^{(ij)}(\rho)
 		={}&
 		(1-p_Z)\rho
 		\nonumber \\
 		&+
 		\frac{p_Z}{3}
 		\left(
 		Z_i\rho Z_i
 		+
 		Z_j\rho Z_j
 		+
 		Z_iZ_j\rho Z_iZ_j
 		\right).
 		\label{eq:two_qubit_phase_flip}
 	\end{align}
 	Here the total phase-flip probability is distributed over the corresponding $Z_i$, $Z_j$, and $Z_iZ_j$ errors.
 	
 	For the asymmetric two-qubit simulations, we again apply the bit-flip and phase-flip channels successively,
 	\begin{align}
 		\mathcal{N}_{XZ}^{(ij)}
 		=
 		\mathcal{N}_{Z}^{(ij)}(p_Z)
 		\circ
 		\mathcal{N}_{X}^{(ij)}(p_X).
 		\label{eq:two_qubit_asymmetric_pauli}
 	\end{align}
 	Here \(\mathcal{N}_{X}^{(ij)}\) and \(\mathcal{N}_{Z}^{(ij)}\) are the two-qubit channels defined in Eqs.~\eqref{eq:two_qubit_bit_flip} and~\eqref{eq:two_qubit_phase_flip}. This composition keeps the total bit-flip and phase-flip probabilities independent while preserving the interpretation of two-qubit errors as errors acting on either one of the two qubits or on both. In contrast to a mutually exclusive Pauli channel, the composed channel also includes mixed \(X\)- and \(Z\)-type two-qubit error events at order \(p_Xp_Z\).
 	
 	The distinction between $X$- and $Z$-type noise is important for \ac{QAOA}. 
 	The cost Hamiltonian in Eq.~\eqref{eq:general_ising} is diagonal in the computational basis and is solely built from Pauli--$Z$ operators, whereas the mixer Hamiltonian in Eq.~\eqref{eq:mixer_hamiltonian} is solely built from Pauli--$X$ operators. 
 	Therefore, bit-flip and phase-flip noise do not generally disturb the \ac{QAOA} evolution in the same way. 
 	This motivates the asymmetric noise sweeps performed in the numerical analysis.
 	
 	\subsection{Gate-Level and Layer-Level Noise Insertion}
 	\label{subsec:gate_layer_noise}
 	
 	The Pauli channels defined in \cref{subsec:pauli_noise_channels} are inserted into the \ac{QAOA} circuit at two different levels. 
 	The noise insertions used in the simulations are indicated by the dashed boxes in \cref{fig:circuits}. 
 	In the layer-level model, the dashed channel acts after a complete cost-and-mixer layer, whereas in the gate-level models it acts after each elementary gate of the chosen decomposition.
 	
 	The first approach is a gate-level noise model, where noise is applied after each elementary gate in the compiled circuit. 
 	The second approach is a layer-level noise model, where an effective noise channel is applied only after a complete \ac{QAOA} layer. 
 	Both models reduce to the ideal \ac{QAOA} circuit when the noise probabilities are set to zero, but they represent different assumptions about how noise accumulates during the computation.
 	
 	In the gate-level model, the Hamiltonian-level \ac{QAOA} layer is first decomposed into elementary gates using one of the gate sets introduced in \cref{subsec:gate_sets}. 
 	A single-qubit Pauli noise channel is applied after each single-qubit gate, and a two-qubit Pauli noise channel is applied after each two-qubit gate. 
 	Thus, for the gate set $\mathcal{G}_1=\{R_z,R_x,\mathrm{CNOT}\}$, noise is inserted after every $R_z$, $R_x$, and CNOT gate. 
 	The $R_z$ and $R_x$ gates are followed by single-qubit channels, while each CNOT gate is followed by a two-qubit channel acting on its control and target qubits.
 	
 	For the gate set $\mathcal{G}_2=\{R_z,R_x,R_{ZZ}\}$, the single-qubit rotations are treated in the same way, but each native $R_{ZZ}$ gate is followed by a two-qubit Pauli channel acting on the two qubits involved in the interaction. 
 	The two gate sets therefore implement the same ideal \ac{QAOA} unitary, but they generally contain different numbers of noisy locations. 
 	In particular, a single $ZZ$ interaction requires two CNOT gates and one $R_z$ gate in $\mathcal{G}_1$, whereas it is implemented by one native $R_{ZZ}$ gate in $\mathcal{G}_2$. 
 	Consequently, the gate-level model is sensitive to the chosen circuit decomposition.
 	
 	In the layer-level model, the \ac{QAOA} layer is treated as a logical cost-and-mixer block. 
 	For layer $\ell$, the ideal operation is
 	\begin{align}
 		U_{\ell}
 		=
 		U_M(\beta_\ell)U_C(\gamma_\ell).
 		\label{eq:qaoa_single_layer_unitary}
 	\end{align}
 	After this complete layer has been applied, an effective Pauli noise channel is applied to the qubits. 
 	This model does not resolve the individual elementary gates inside $U_C(\gamma_\ell)$ and $U_M(\beta_\ell)$. 
 	Instead, it represents the accumulated effect of noise over one full \ac{QAOA} layer in a coarse-grained way.
 	
 	The noisy final state obtained from either noise-insertion model is denoted by $\rho_d^{\mathrm{noisy}} $. The corresponding noisy cost value is evaluated as
 	\begin{align}
 		C_d^{\mathrm{noisy}}(\boldsymbol{\gamma},\boldsymbol{\beta})
 		=
 		\operatorname{Tr}
 		\left[
 		H_C\rho_d^{\mathrm{noisy}}
 		\right].
 		\label{eq:noisy_qaoa_cost}
 	\end{align}
 	This is the quantity used to compare the effect of bit-flip and phase-flip errors on the optimized \ac{QAOA} performance.
 	
 	The distinction between the two noise-insertion models is important for the interpretation of the results. 
 	Gate-level noise resolves the compiled circuit and therefore depends on the number of elementary gates used by a particular decomposition. 
 	Layer-level noise instead abstracts away the compilation details and treats each \ac{QAOA} layer as a single noisy logical block. 
 	By comparing both descriptions, we can determine whether an observed asymmetry between bit-flip and phase-flip noise is a robust feature of the \ac{QAOA} dynamics or a consequence of the chosen gate decomposition.

 	\section{Noise Simulation}
 	\label{sec:noise_simulation}
 	
 	After defining the \ac{QAOA} circuit decompositions and the noise channels, we now describe the numerical protocol used to simulate noisy \ac{QAOA}. 
 	The \ac{FGA} instance was first formulated as a \ac{QUBO} and mapped to an Ising Hamiltonian using Quark~\cite{quark2024}. 
 	PennyLane was used for Hamiltonian handling and for constructing the transverse-field mixer Hamiltonian~\cite{pennylane}. 
 	The corresponding \ac{QAOA} circuits and noise models were then implemented in Qiskit, and the noisy dynamics were simulated using exact density-matrix evolution with Qiskit Aer~\cite{qiskit}. 
 	The purpose of this section is to make clear how the noisy circuits are evaluated, how the variational parameters are optimized, and how the final performance is compared across noise models, gate sets, and \ac{QAOA} depths.
 	
 	\subsection{Simulation Parameters}
 	\label{subsec:simulation_parameters}
 	
 	All simulations are performed on the six-qubit \ac{FGA} instance introduced in \cref{subsec:ising_mapping}. 
 	For each \ac{QAOA} depth $d$, the circuit is parameterized by
 	\begin{align}
 		\theta
 		=
 		(\gamma_1,\ldots,\gamma_d,\beta_1,\ldots,\beta_d),
 	\end{align}
 	where the $\gamma_\ell$ parameters control the cost unitaries and the $\beta_\ell$ parameters control the mixer unitaries. 
 	For a fixed value of $\theta$, the circuit prepares either the ideal state $\rho_d$ or the noisy state $\rho_d^{\mathrm{noisy}}$, depending on whether the noise probabilities are set to zero or to finite values.
 	
 	The cost value is evaluated exactly from the final density matrix,
 	\begin{align}
 		C_d(\theta)
 		=
 		\operatorname{Tr}
 		\left[
 		H_C \rho_d
 		\right],
 		\qquad
 		C_d^{\mathrm{noisy}}(\theta)
 		=
 		\operatorname{Tr}
 		\left[
 		H_C \rho_d^{\mathrm{noisy}}
 		\right].
 	\end{align}
 	Thus, the simulations do not include finite-shot sampling noise. 
 	This choice is intentional as it isolates the effect of the physical noise channels from statistical fluctuations due to measurement sampling. 
 	Consequently, any degradation in the optimized cost can be attributed to the applied bit-flip and phase-flip channels rather than to finite measurement statistics. 
 	
 	As the \ac{QUBO} problem is non-convex, a global optimum is not guaranteed to be found and the optimized result will depend on the initial parameter vector. 
 	To reduce this initialization bias, we repeat the optimization from several independent random initial points. 
 	For the $r$-th iteration, the initial parameter vector is sampled as
 	\begin{align}
 		\theta^{(0,r)}
 		\sim
 		\mathrm{Uniform}\left([0,2\pi]^{2d}\right),
 		\qquad
 		r=1,\ldots,N_{\mathrm{seed}} .
 	\end{align}
 	Each iteration is optimized independently using the same classical optimizer and the same stopping criteria. 
 	This gives a set of locally optimized parameter vectors
 	\begin{align}
 		\theta_{\mathrm{opt}}^{(r)}
 		=
 		\arg\min_{\theta}
 		C_d^{\mathrm{noisy}}(\theta),
 		\qquad
 		r=1,\ldots,N_{\mathrm{seed}},
 	\end{align}
 	for the corresponding noisy circuit setting. 
 	The optimized cost obtained from iteration $r$ is denoted by
 	\begin{align}
 		C_{\mathrm{opt}}^{(r)}
 		=
 		C_d^{\mathrm{noisy}}
 		\left(
 		\theta_{\mathrm{opt}}^{(r)}
 		\right).
 	\end{align}
 	
 	From the set of optimized restart values, we extract the best and worst optimized costs
 	\begin{align}
 		C_{\mathrm{best}}
 		=
 		\min_r C_{\mathrm{opt}}^{(r)},
 		\text{ and }
 		C_{\mathrm{worst}}
 		=
 		\max_r C_{\mathrm{opt}}^{(r)} .
 		\label{eq:best_worst_cost}
 	\end{align}
 	
 	
 	The best value $C_{\mathrm{best}}$ is used to characterize
 	the lowest cost obtained after classical optimization.
 	Since our primary objective is to investigate how
 	different noise channels affect the optimized QAOA
 	performance, we are interested in the best solution
 	obtained from multiple random initializations, rather
 	than the outcome of a particular optimization run. This choice reduces the dependence of the comparison
 	on unfavorable parameter initializations and allows
 	us to focus on the effect of the noise channels.
 	However, $C_{\mathrm{best}}$ characterizes the best
 	result found for a fixed number of optimization runs
 	and does not necessarily represent the performance
 	of a typical individual run.

 	The interval between $C_{\mathrm{best}}$ and
 	$C_{\mathrm{worst}}$ is used in the plots as a shaded
 	region, indicating the spread caused by different
 	random initializations. This range illustrates the variability of the
 	optimization outcomes, although it should not be
 	interpreted as a statistical confidence interval.
 	
 	The number of random initializations and the
 	classical optimization procedure are kept fixed
 	across all simulation settings. The \ac{QAOA} parameters were optimized using the derivative-free COBYLA algorithm
 	as implemented in \texttt{scipy.optimize.minimize}. Since COBYLA is not gradient-based, no learning rate was used and no gradient or parameter-shift estimation was performed. The objective function was evaluated exactly from the final density matrix, without finite-shot sampling noise. For each \ac{QAOA} depth \(d\) and each noise setting, we used \(N_{\mathrm{seed}}=50\) independent random initializations sampled uniformly from \([0,2\pi]^{2d}\), with random seed \(123\). No explicit parameter bounds were imposed during the optimization. The COBYLA initial trust-region radius was set to \(0.5\), the
 	optimizer tolerance was \(10^{-5}\), and the maximum number of objective-function evaluations was \(300\) per restart. We also used an early-stopping criterion: if the best objective value did not improve by more than \(10^{-5}\) for \(30\) consecutive objective-function evaluations, the corresponding restart was terminated.

 	We perform this protocol for the ideal circuit and for noisy circuits under both gate-level and layer-level noise models. 
 	For gate-level simulations, the circuit is compiled using either $\mathcal{G}_1=\{R_z,R_x,\mathrm{CNOT}\}$ or $\mathcal{G}_2=\{R_z,R_x,R_{ZZ}\}$, and Pauli noise is inserted after the elementary gates. 
 	For layer-level simulations, an effective Pauli channel is inserted after each complete \ac{QAOA} layer. 
 	The same cost Hamiltonian, initialization rule, and optimization protocol are used across all cases, so that differences in the final performance can be attributed to the noise model, the noise probabilities, and the chosen circuit decomposition.

 	\subsection{Performance Metric}
 	\label{subsec:performance_metric}
 	
 	To compare the performance of different noisy \ac{QAOA} simulations, we measure how far the optimized cost is from the exact ground-state energy of the cost Hamiltonian. 
 	Since the \ac{FGA} instance considered here contains only six qubits, the minimum eigenvalue of the cost Hamiltonian $H_C$, including the constant offset $\alpha$, is computed exactly by diagonalization. 
 	We denote this value by $\lambda_{\min}$. This value represents the optimal cost encoded by the Ising Hamiltonian.
 	
 	For each simulation setting, the main performance metric is defined as
 	\begin{align}
 		\Delta
 		=
 		C_{\mathrm{best}}
 		-
 		\lambda_{\min},
 		\label{eq:performance_delta}
 	\end{align}
 	where $C_{\mathrm{best}}$ is the best optimized cost obtained over all iterations, as defined in Eq.~\eqref{eq:best_worst_cost}. 
 	The quantity $\Delta$ therefore measures the residual error of the optimized noisy \ac{QAOA} circuit relative to the exact optimum. 
 	A smaller value of $\Delta$ indicates better performance, while $\Delta=0$ would correspond to reaching the exact ground-state energy.
 	It also allows direct comparison between the ideal circuit and the noisy circuits. In the absence of noise, $\Delta$ measures the limitation caused by finite \ac{QAOA} depth and the classical optimization procedure. 
 	In the noisy case, additional increases in $\Delta$ quantify the degradation caused by the applied noise channels.
 	
 	For each noise setting, we report $\Delta$ using the best optimized cost over the all iterations. 
 	In addition, the spread between the best and worst optimized costs is shown as a shaded region in the corresponding plots. It reflects the sensitivity of the optimization outcome to the random initialization of the \ac{QAOA} parameters.
 	
 	Using $\Delta$ as the main metric allows us to compare bit-flip and phase-flip noise on the same scale. 
 	For asymmetric noise sweeps, we evaluate
 	\begin{align}
 		\Delta(p_X,p_Z)
 		=
 		C_{\mathrm{best}}(p_X,p_Z)
 		-
 		\lambda_{\min},
 	\end{align}
 	where $p_X$ and $p_Z$ are varied independently. 
 	The resulting values show whether the \ac{QAOA} circuit is more sensitive to one type of Pauli error than to the other.

 	\subsection{Results}
 	\label{subsec:result}
 	
 	\begin{figure*}[tbph]
 		\centering
 		\includegraphics[width=\linewidth]{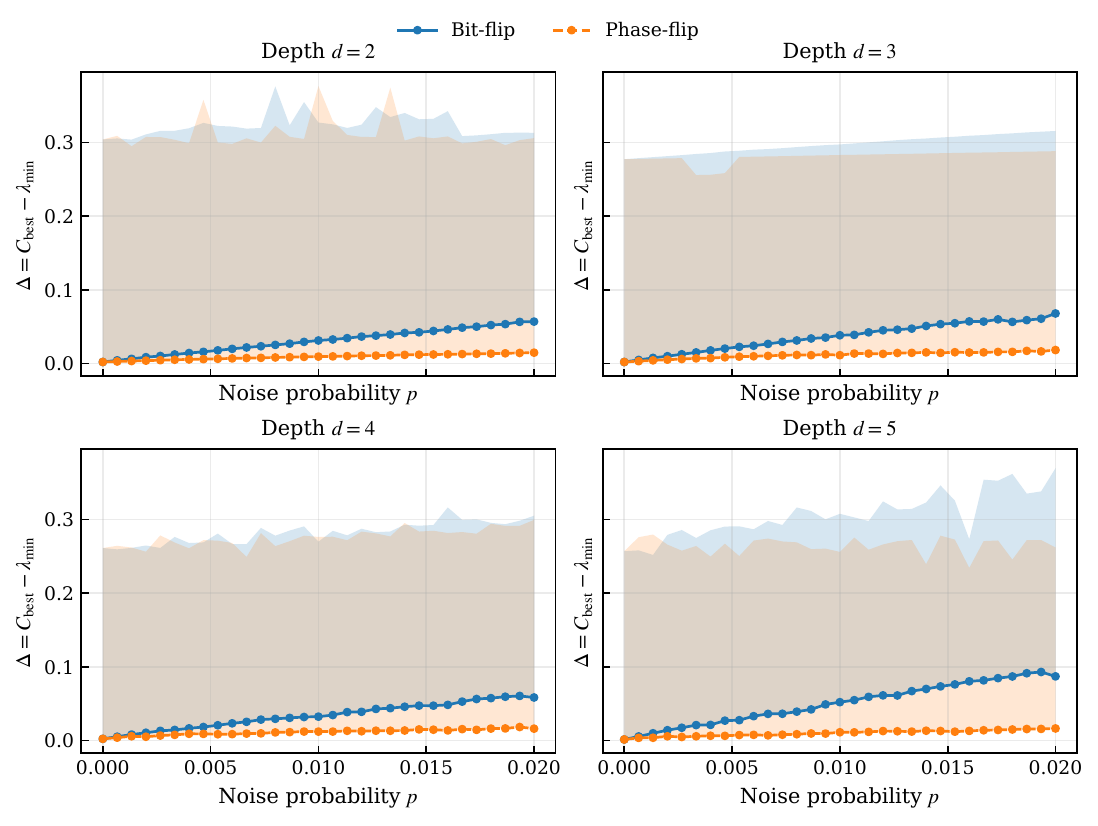}
 		\caption{The deviation $\Delta$ from the ideal cost $C_{\mathrm{cost}}$ for different \ac{QAOA} depths as a function of the physical noise parameter $p$. The shaded area indicates the runs from different initial parameters, while the circles correspond to their minimum. Lines are just there to guide the eye. This simulation uses the circuit at the layer level, i.e. without compilation to basic gates.}
 		\label{fig:layer_1D}
 	\end{figure*}
 	
 	\begin{figure*}[tbph]
 		\centering
 		\includegraphics[width=\linewidth]{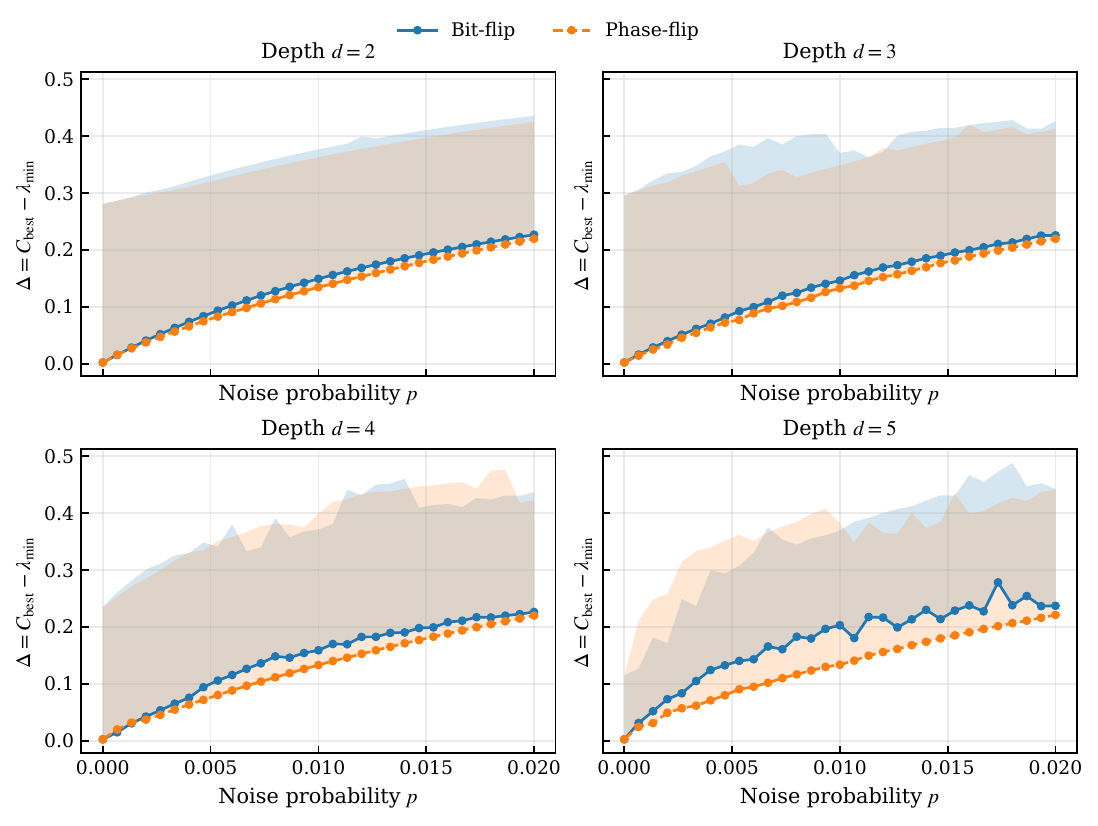}
 		\caption{The deviation $\Delta$ from the ideal cost $C_{\mathrm{cost}}$ for different \ac{QAOA} depths as a function of the physical noise parameter $p$. The shaded area indicates the runs from different initial parameters, while the circles correspond to their minimum. Lines are just there to guide the eye. This simulation uses the circuit compiled to \ac{CNOT}, $R_{X}$, and $R_Z$ gates.}
 		\label{fig:gate_decomposed_1D}
 	\end{figure*}
 	
 	\begin{figure*}[tbph]
 		\centering
 		\includegraphics[width=\linewidth]{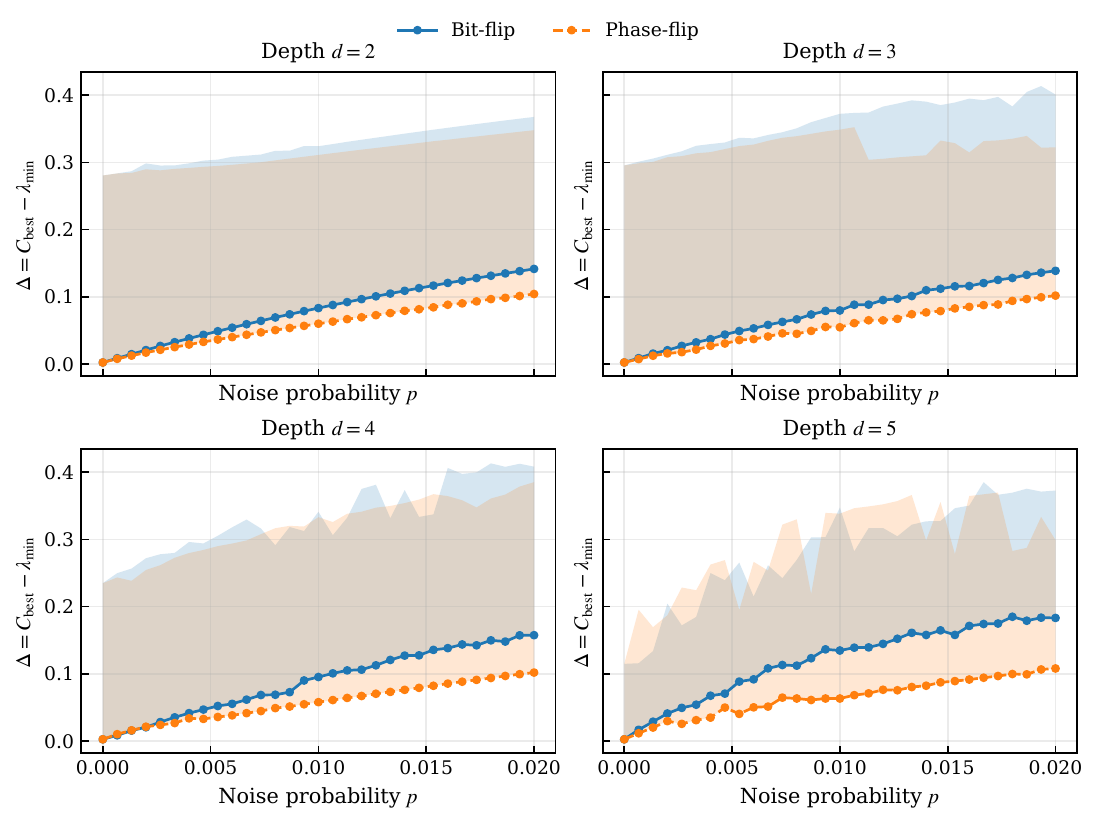}
 		\caption{The deviation $\Delta$ from the ideal cost $C_{\mathrm{cost}}$ for different \ac{QAOA} depths as a function of the physical noise parameter $p$. The shaded area indicates the runs from different initial parameters, while the circles correspond to their minimum. Lines are just there to guide the eye. This simulation uses the circuit compiled to $R_{ZZ}$, $R_{X}$, and $R_{Z}$ gates.}
 		\label{fig:gate_native_1D}
 	\end{figure*}
 	
 	\begin{figure*}[tbph]
 		\centering
 		\includegraphics[width=\linewidth]{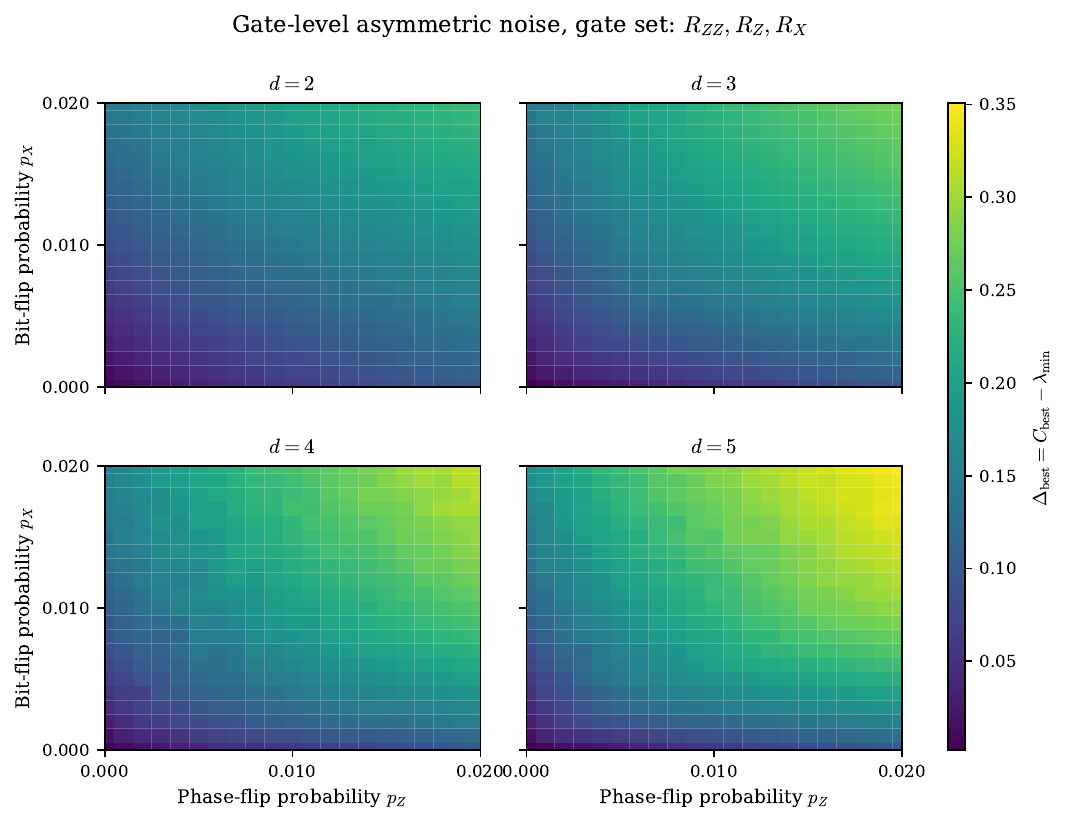}
 		\caption{
 			Gate-level asymmetric noise heatmap for the QAOA circuit compiled into the gate set
 			\(\mathcal{G}_2=\{R_Z,R_X,R_{ZZ}\}\).
 			Each panel shows the optimized residual cost
 			\(\Delta_{\mathrm{best}}=C_{\mathrm{best}}-\lambda_{\min}\) for depths
 			\(d=2,3,4,5\), as a function of the independently varied bit-flip and phase-flip
 			probabilities \(p_X\) and \(p_Z\). The same color scale is used for all depths
 			within the figure. Compared with the CNOT-based decomposition, this native
 			\(R_{ZZ}\) implementation contains fewer two-qubit noisy locations for each
 			Ising interaction, so differences between the two heatmaps reflect both the
 			chosen gate decomposition and the asymmetric Pauli noise response.
 		}
 		\label{fig:gate_heatmap_native}
 	\end{figure*}
 	
 	\begin{figure*}[tbph]
 		\centering
 		\includegraphics[width=\linewidth]{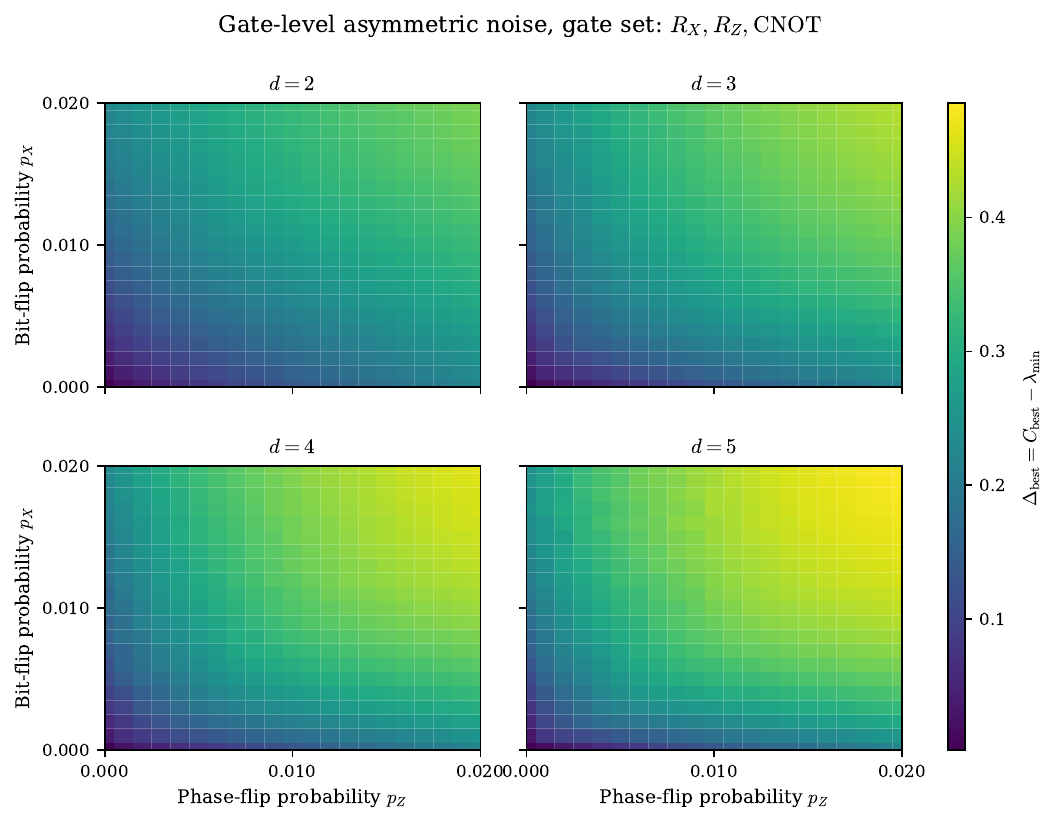}
 		\caption{
 			Gate-level asymmetric noise heatmap for the QAOA circuit compiled into the gate set
 			\(\mathcal{G}_1=\{R_Z,R_X,\mathrm{CNOT}\}\).
 			Each panel corresponds to a different QAOA depth \(d=2,3,4,5\).
 			The horizontal axis shows the phase-flip probability \(p_Z\), while the vertical
 			axis shows the bit-flip probability \(p_X\). The color scale represents the
 			optimized residual cost
 			\(\Delta_{\mathrm{best}}=C_{\mathrm{best}}-\lambda_{\min}\), where
 			\(C_{\mathrm{best}}\) is the lowest optimized cost obtained over
 			\(N_{\mathrm{seed}}=50\) random initializations. Smaller values of
 			\(\Delta_{\mathrm{best}}\) correspond to better agreement with the exact
 			ground-state energy. Noise is applied after the elementary gates of the compiled
 			circuit.
 		}
 		\label{fig:gate_heatmap_decomposed}
 	\end{figure*}
 	
 	\begin{figure*}[tbph]
 		\centering
 		\includegraphics[width=\linewidth]{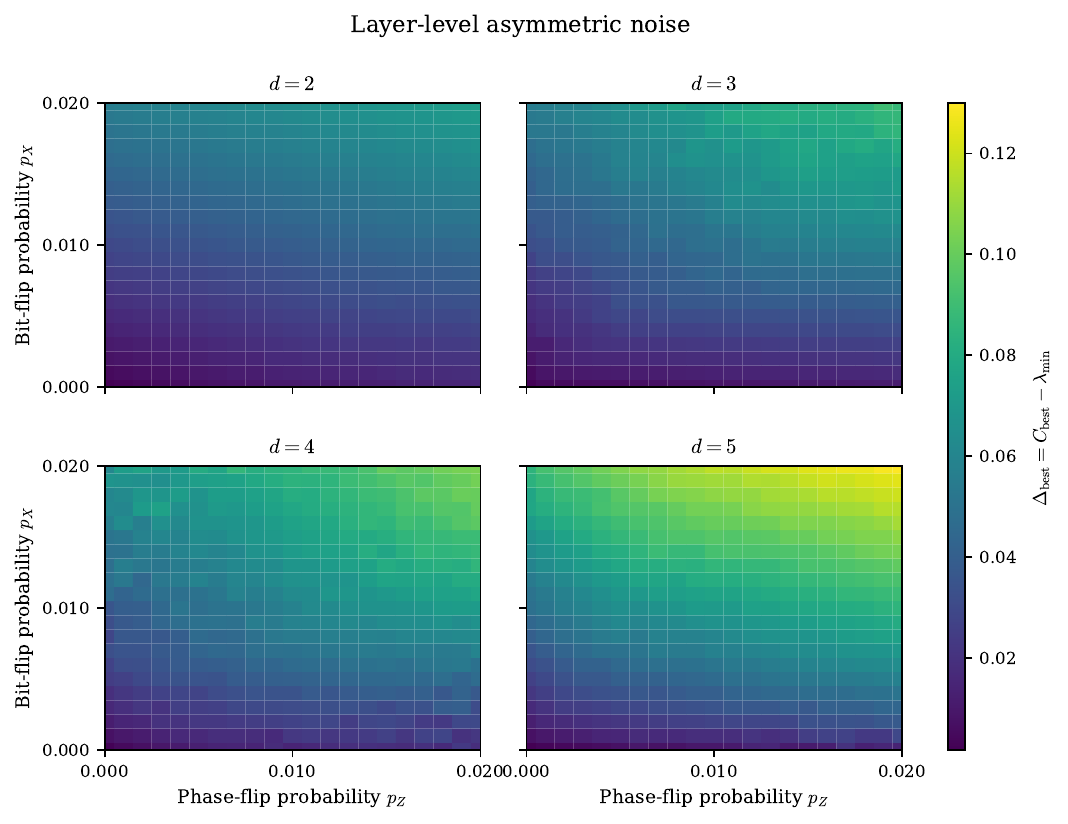}
 		\caption{
 			Layer-level asymmetric noise heatmap for the QAOA circuit.
 			In this model, the cost and mixer operations of one QAOA layer are treated as a
 			single logical block, and the effective Pauli noise channel is applied after each
 			complete layer. The horizontal axis denotes the phase-flip probability \(p_Z\),
 			the vertical axis denotes the bit-flip probability \(p_X\), and the color scale
 			shows
 			\(\Delta_{\mathrm{best}}=C_{\mathrm{best}}-\lambda_{\min}\).
 			The panels correspond to QAOA depths \(d=2,3,4,5\). Since this model does not
 			resolve the elementary gates inside the layer, it provides a coarse-grained
 			comparison with the gate-level simulations.
 		}
 		\label{fig:layer_heatmap}
 	\end{figure*}
 	We now compare the optimized \ac{QAOA} performance under bit-flip and phase-flip
 	noise for the six-qubit \ac{FGA} instance. 
 	The performance is quantified by the
 	residual cost $\Delta$ as defined in  \cref{{eq:performance_delta}}. 
 	Smaller values of  $\Delta$ correspond to better optimized QAOA performance. 
 	In \cref{fig:layer_1D,fig:gate_decomposed_1D,fig:gate_native_1D}, $\Delta$ is shown as a function of the noise
 	probability $p$ for the two limiting cases $p_X=p, p_Z=0$ and
 	$p_X=0, p_Z=p$. The heatmaps in
 	\cref{fig:gate_heatmap_native,fig:gate_heatmap_decomposed,fig:layer_heatmap} show the combined asymmetric case in which $p_X$ and
 	$p_Z$ are varied independently.  
 	We discuss these results in more detail in the following paragraphs and start with two observations regarding how noise affects the cost-function.
 	
 	On the one hand, there is a contribution from error propagation due to the structure of the \ac{QAOA} circuit.
 	The cost evolution $U_C$ is generated by Pauli-\(Z\) and Pauli-\(ZZ\) terms, while
 	the mixer evolution $U_M$ is generated by Pauli-\(X\) terms. Therefore, \(Z\)-type
 	phase-flip errors commute with the ideal cost evolution for arbitrary rotation
 	angles
 	but not with $U_M$.
 	In contrast $X$-errors (bit-flips) commute with the mixer $U_M$ but not with cost evolution $U_C$.
 	As multi-qubit interactions only occur in $U_C$, only bit-flip errors can spread inside the layer-level circuit.
 	Note that neither error type commutes with the complete \ac{QAOA} circuit,
 	because the algorithm alternates between the non-commuting cost and mixer evolutions. 
 	
 	On the other hand, we note which errors affect the application at the measurements.
 	As the cost
 	Hamiltonian is diagonal in the computational basis,
 	i.e. we perform a computational basis measurement at the end of the circuit,
 	only bit-flip errors affect the measurement.
 	They directly transfer population between different assignment states and
 	immediately change the cost expectation value. 
 	Phase-flip errors, in contrast, have no effect at the stage of the measurements.
 	Their effect becomes relevant mainly through subsequent mixer rotations, which
 	do not commute with \(Z\)-type errors. 
 	
 	When considering noise at the layer-level, see \cref{fig:circuits_layers},
 	the corresponding Pauli channel is applied only once
 	after each complete \ac{QAOA} layer in contrast to after every elementary gate in the other two circuits.
 	Therefore,  
 	errors are inserted much less frequently
 	compared to the gate-level simulations.
 	This
 	explains why the residual cost remains smallest in the layer-level model for
 	both bit-flip and phase-flip noise, see \cref{fig:layer_1D}. 
 	For all depths considered, $\Delta$ increases with the
 	noise probability, but the increase is considerably stronger for bit-flip noise
 	than for phase-flip noise.
 	Thus, the layer-level results already indicate that bit-flip
 	noise is the more severe error channel for the optimized cost.
 	This can be understood as in the layer-level simulation the effect  of error propagation remains weak, so the optimized residual cost under phase-flip
 	noise stays close to the noiseless value over the investigated range of noise
 	probabilities. 
 	
 	The deviation from the ideal performance $\Delta$ for the circuits with gate-level noise as a function of the bit-flip or phase-flip error rate are shown in
 	\cref{fig:gate_decomposed_1D,fig:gate_native_1D}. 
 	In these
 	simulations, the errors occur with the same probabilities, but they are inserted after every elementary gate in the compiled circuit. The
 	gate-level model therefore contains many more noisy locations than the
 	layer-level model, and the optimized residual cost is correspondingly larger.
 	It also makes the result sensitive to the chosen gate decomposition, because
 	different decompositions implement the same ideal QAOA unitary with different
 	numbers of single- and two-qubit gates.

 	It is important to note that the layer-level and
 	gate-level noise models do not correspond to the same
 	total accumulated noise, as the number and
 	placement of noisy operations differ between the models
 	and circuit decompositions. 
 	Consequently, the smaller
 	residual costs obtained in the layer-level simulations
 	should not be interpreted as an intrinsic advantage
 	of this noise model. 
 	Our purpose in considering these different implementations
 	is therefore not to compare their absolute performance,
 	but to investigate how the relative sensitivity to
 	bit-flip and phase-flip errors changes with the
 	noise model and circuit decomposition.
 	
 	For the CNOT-based decomposition with gate set
 	\mbox{\(\mathcal{G}_1=\{R_Z,R_X,\mathrm{CNOT}\}\)}, shown in
 	\cref{fig:gate_decomposed_1D}, both noise channels lead to a clear increase in \(\Delta\). 
 	Because each \(ZZ\) interaction is implemented through a
 	\(\mathrm{CNOT}\)-\(R_Z\)-\(\mathrm{CNOT}\) sequence,
 	this has more elementary gates, in particular two-qubit gates, and thus more error locations.
 	The bit-flip curve remains above the phase-flip curve, confirming
 	that bit-flip noise is still the more damaging channel. 
 	However, the separation
 	between the two curves is less pronounced than in the native-\(R_{ZZ}\) circuit.
 	One reason for this is that for CNOT gates, in contrast to the $R_{ZZ}$-gate, phase-flip errors can propagate from the target to the control qubit.
 	Thus, the CNOT decomposition does not remove the dominance of bit-flip noise,
 	but it makes phase-flip noise significant as well.
 	
 	The native-\(R_{ZZ}\) gate-level results are shown in
 	\cref{fig:gate_native_1D}. In this implementation, each two-qubit Ising
 	interaction is represented directly by one \(R_{ZZ}\) gate instead of a longer CNOT-based sequence. 
 	The number of noisy two-qubit locations is therefore reduced. 
 	Since \(Z\)-type phase-flip errors commute with the \(R_Z\) and
 	\(R_{ZZ}\) cost gates, their effect remains mainly limited to the subsequent
 	non-commuting \(R_X\) mixer rotations. Bit-flip errors, in contrast, do not
 	commute with the \(Z\)-structured cost operations and can directly change the
 	computational-basis assignment states. For this reason, the native-\(R_{ZZ}\)
 	results show the most pronounced asymmetry between the two noise channels:
 	bit-flip noise produces a clearly larger increase in \(\Delta\) than phase-flip
 	noise, especially at larger depths and larger noise probabilities.
 	
 	The same qualitative picture emerges when bit-flip and phase-flip noise are applied simultaneously, see heat maps in \cref{fig:gate_heatmap_native,fig:gate_heatmap_decomposed,fig:layer_heatmap}. 
 	As expected, the
 	optimized residual cost \(\Delta_{\mathrm{best}}\) is smallest near the
 	noiseless point \((p_X,p_Z)=(0,0)\) and increases as either error probability is
 	raised in all three models. The layer-level heatmap in \cref{fig:layer_heatmap} has the
 	smallest overall scale, consistent with the smaller number of noisy insertions
 	in this model. The gate-level heatmaps in
 	\cref{fig:gate_heatmap_native,fig:gate_heatmap_decomposed} show
 	larger residual costs because noise is applied after each elementary gate.
 	Comparing the two gate-level heatmaps again shows that the native-\(R_{ZZ}\)
 	circuit displays the bit-flip dominance more clearly, while the CNOT-based
 	decomposition increases the accumulated noise and makes phase-flip errors more
 	visible. Thus, the heatmaps support the conclusion drawn from the
 	one-dimensional sweeps: bit-flip noise is the most severe error channel in our
 	analysis, while the relative importance of phase-flip noise is enhanced when
 	the circuit decomposition introduces additional noisy gates.
 	
 	The comparison is performed using the same cost
 	Hamiltonian and fixed penalty coefficients throughout
 	the simulations. While the relative strengths of
 	the penalty terms can influence the energy landscape
 	and, consequently, the sensitivity to different
 	error channels, they are not varied in the present
 	analysis. Therefore, our results characterize the
 	relative effects of bit-flip and phase-flip noise
 	for the considered penalty setting. A systematic
 	investigation of the dependence on these coefficients
 	would provide further insight into the robustness
 	of the observed noise asymmetry.
 	
 	Taken together, the one-dimensional sweeps and the heatmaps show that bit-flip
 	and phase-flip noise do not affect the optimized QAOA cost equally. For the
 	FGA instance considered here, bit-flip noise consistently produces the larger
 	degradation in \(\Delta\). The asymmetry is clearest in the layer-level and
 	native-\(R_{ZZ}\) simulations. In the CNOT-based gate-level model, the
 	additional elementary gates introduce more noisy locations and therefore make
 	the phase-flip contribution significant as well, but bit-flip noise remains the
 	dominant source of degradation. These results motivate the use of asymmetric
 	error-correction strategies. The error type that is most damaging for the
 	application-level objective should receive stronger protection than an error
 	type with a weaker effect on the optimized cost.

 	\section{Tailored Asymmetrical Quantum Error Correction}
 	\label{sec:error_correction}
 	The numerical results in \cref{subsec:result} show that bit-flip and phase-flip noise do not affect the \ac{QAOA} output equally. 
 	For the \ac{FGA} instance considered here, the optimized residual cost increases more strongly under bit-flip noise than under phase-flip noise.
 	The observed asymmetry can be understood from structural
 	properties of the QAOA circuit that are not specific
 	to the considered FGA instance. For the standard QAOA
 	ansatz with a diagonal Ising cost Hamiltonian and a
 	transverse-field mixer, $Z$-type errors commute with
 	the cost evolution, whereas $X$-type errors commute
 	with the mixer evolution. Furthermore, bit-flip errors
 	can directly affect the computational-basis measurement
 	outcomes, while phase-flip errors occurring immediately
 	before measurement do not. These properties, together with the consistent
 	qualitative behavior observed across different QAOA
 	depths, noise models, and circuit decompositions,
 	provide a physical basis for expecting similar
 	noise asymmetries in other Ising-based QAOA
 	applications. Nevertheless, the relative sensitivity
 	to the two error channels may depend on the particular
 	problem Hamiltonian, penalty coefficients, and circuit
 	implementation.

 	These observations motivate the use of QEC strategies
 	tailored to the application-level impact of different
 	noise channels. In particular, such a strategy may be
 	more resource-efficient than one that protects equally
 	against all Pauli errors. If one error type has a
 	particularly severe effect on the application-level
 	performance, more of the available error-correction
 	resources can be allocated to suppressing that error type.

 	
 	The efficient use of resources for error suppression is particularly important in the early fault-tolerant regime, because full quantum error correction can require a large number of additional physical qubits and operations for syndrome-extraction, making it difficult to implement on near-term devices~\cite{Gaitan2009,Lidar_Brun_2013}. 
 	A symmetric code that protects equally against bit-flip and phase-flip errors can therefore be unnecessarily expensive if the algorithm is mainly limited by only one of these errors. 
 	In such a situation, an asymmetric code can be more resource-efficient. 
 	It can provide stronger protection against the dominant or more damaging error channel, while using less protection against the less relevant one. 
 	
 	Asymmetric quantum error correction codes have been developed to tackle asymmetric noise~\cite{Steane96b,ioffe2007asymmetric}.
 	Here we employ them for a different reason: The noise might be symmetric, but its effect on the performance of the application is different. More precisely, physical noise bias refers to unequal bit-flip and phase-flip error probabilities ($p_X\neq p_Z$), whereas algorithmic noise sensitivity describes their different effects on the optimized cost, even for equal physical error probabilities. QEC code asymmetry, on the other hand, refers to unequal protection against the two error types.
 	
 	Hence the error that should be corrected most strongly is not always the one with the largest probability to occur physically. 
 	Instead, it is the one that produces the largest degradation in the final objective value. 
 	In our analysis, the choice of an asymmetric code is guided by how strongly the metric $\Delta$ defined in  \cref{eq:performance_delta} increases under each noise channel, rather than by the physical noise probabilities alone. 
 	
\begin{figure}[tbph]
 		\centering
 		\def\svgwidth{\linewidth}
 		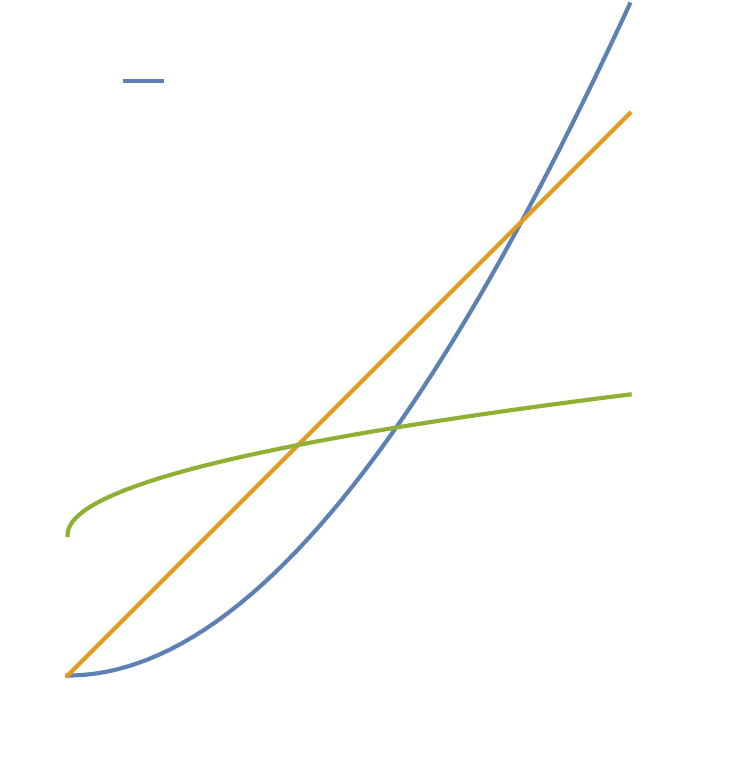
 		\caption{
 			Simplified effective-noise model for application-level projection. When using a quantum error correction code, the physical error rate is replaced by an effective logical error rate. The figure shows the physical error probability on one axis and the residual logical error probability on the other. Without error correction, both are equal. With error correction, each error probability is shifted to a lower effective error probability, so that the same physical noise level produces a smaller logical error. The corresponding projected improvement in terms of deviation from the minimal cost is shown on the right axis. This figure represents an effective-noise projection and not a complete fault-tolerant QEC circuit simulation.
 		}
 		\label{fig:fromptopL}
 	\end{figure}

 	Before discussing a specific asymmetric code, it is useful to recall the general role of quantum error correction. 
 	At an abstract level, a quantum error-correcting code maps a physical error probability $p$ to a residual logical error $p_L$. 
 	If the code is effective in the relevant noise regime, then the logical error probability is smaller than the physical one. 
 	Thus, the main effect of error correction can be viewed as replacing the physical noise strength by an effective logical noise strength. 
 	This idea is illustrated schematically in \cref{fig:fromptopL}.
 	Effectively this reduces the cost $C(p)$ as a function of the physical noise $p$ to $C(p_L(p))$.
 	While this is not a full simulation of the quantum error-corrected circuit, this model allows us to analyze the effect of the asymmetric error correction code specific to our application without rerunning the simulation for $C$.

 	We now illustrate this idea with an asymmetric \ac{CSS} code~\cite{Calderbank96,Steane96a} which is a generalization of the 9-Qubit Shor code~\cite{Shor95}, sometimes called the \ac{QPC}~\cite{Hayes08}.
 	It is the concatenation of an inner repetition code of length $n_2$ and an outer repetition code of length $n_1$ to protect against bit-flip and phase-flip errors, respectively.
 	In total the code uses $N = n_1 n_2$ physical qubits to encode one logical qubit. 
 	Better error correction codes are known, but we choose the \ac{QPC} for clearly illustrating the effect of the code's asymmetry.
 	In the following $n_1$ and $n_2$ are odd numbers.
 	
 	A logical $X$ error occurs, if a majority of the inner blocks introduce $Z$-errors.
 	A logical $Z$ error occurs, if an odd number of inner blocks fail, which happens if the majority of qubits of a block has an $X$ error.
 	Thus, the logical error rates read
 	\begin{alignat}{2}
 		p_x^{(L)} =& P_{\mathrm{maj}}(P_{\mathrm{odd}}(p_z, n_2), n_1)  \\
 		\text{and }   p_z^{(L)} =& P_{\mathrm{odd}}(P_{\mathrm{maj}}(p_x, n_2), n_1),
 	\end{alignat}
 	where
 	\begin{alignat}{2}
 		P_{\mathrm{maj}}(p, n) =& \sum_{k=\left\lfloor\frac{n}{2} \right\rfloor+1} \left(\begin{array}{c}
 			n  \\
 			k 
 		\end{array}\right) p^k (1-p)^{n-k}\\
 		\text{and }    P_{\mathrm{odd}}(p, n) =& \frac{1}{2}(1-(1-2p)^n)
 	\end{alignat}
 	and $p_X$ and $p_Z$ are, like in our noise model of \cref{sec:noise_modeling}, the probabilities that a physical bit- or phase-flip error occurred, respectively.
 	
\begin{figure}[tbph]
 		\centering
 		\def\svgwidth{\linewidth}
 		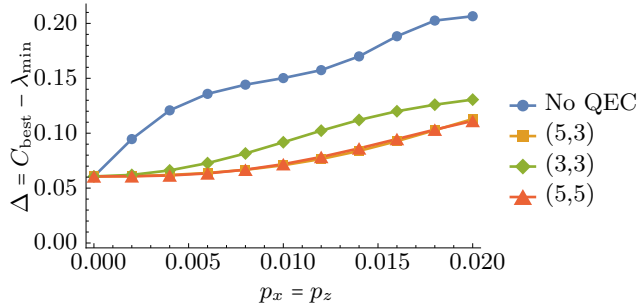
 		\caption{The performance of the QAOA algorithm when using different sizes of the outer and inner repetition code of the \ac{QPC}, $(n_1, n_2)$. The lower $\Delta$ the better. The shown plot corresponds to gate level noise with native $R_{\mathrm{ZZ}}$ gates for $d=2$ layers and similar plots can be drawn for the other cases considered above. The \mbox{(5, 5)} code beats the standard 9-qubit Shor code \mbox{(3, 3)}, but due to the asymmetric effect of the noise on the cost function, the \mbox{(5, 3)} code suffices to achieve the same results. Lines only guide the eye. The visible numerical inaccuracies are inherited from our numerical simulation, see \cref{subsec:result}. Instead of simulating the full error correction circuit, the performance is estimated from replacing physical error rates by effective logical error rates, see main text for details.}
 		\label{fig:QPC53}
 	\end{figure}

 	We now, separately for every $d\in\{2,3,4,5\}$ and every noise model, fit a polynomial to the minimal $\Delta$ of our simulation data presented above and replace the physical error rates by the logical error rates.
 	We use polynomials of degree five both in $p_x$ and $p_z$, see the supplementary material for the explicit expressions.
 	The goodness of the fits expressed as the coefficient of determination $R^2$ are $99.995 \%$, $99.942 \%$, $99.290 \%$ and $99.677 \%$ for $d=2, 3, 4,$ and $5$, respectively.
 	The fit allows us to estimate how \ac{QEC} codes affect the performance of our application algorithm.
 	The idea is to employ an asymmetric error correction code to tackle bit flip errors which proved to be more severe to our application.
 	In terms of the \ac{QPC} we expect $n_1$ to be more important than $n_2$, because $p^{(L)}_X$ affects the application more severely than $p^{(L)}_Z$ and it is reduced by the majority vote of size $n_1$.
 	And indeed choosing $n_1=5$ and $n_2=3$ works as well as using $n_1=n_2=5$, with only $N=15$ instead of $25$ qubits.
 	The resulting $\Delta$ is shown in \cref{fig:QPC53}.
 	
 	When elevating the physical QAOA circuit to a logical, error-corrected, quantum circuit, some overhead is added.
 	In the early fault-tolerant regime targeted by this work, the circuit gadgets used for syndrome extraction add considerable noise.
 	Ideally we would like to take this into account when choosing the asymmetry of the employed error correction code.
 	This can be done via the calibration method introduced in~\cite{Wimmer2024}.
 	With as little overhead as a single additional calibration experiment for the syndrome extraction circuit, this method allows to get multiplicative factors to the expectation values of elements of the stabilizer of the code space.
 	The larger these factors deviate from one, the stronger the corresponding expectation value is affected by the noise.
 	For stabilizer elements that contain only $X$ or $Z$ operators, this indicates how many $Z$ or $X$ errors are introduced, and
 	the asymmetry of the code should be tuned towards better error correction capabilities for phase flip and bit flip errors,  respectively.
 	This approach is complementary to the one discussed above, as it does not take into account the application specific aspects but allows for estimation and adjustments at runtime (in contrast to the simulations presented here).
 	More research on the combination of these two methods to estimate and mitigate noise seems to be promising.
 	
 	\section{Conclusion}
 	\label{sec:conclusion}
 	In this work, we studied how bit-flip and phase-flip noise affect the
 	performance of \ac{QAOA} for a six-qubit flight-gate assignment instance. The
 	problem was formulated as a \ac{QUBO}, mapped to an Ising cost Hamiltonian, and
 	simulated with exact density-matrix evolution.
 	The noisy performance was
 	measured using the residual cost
 	$ \Delta=C_{\mathrm{best}}-\lambda_{\min}$, where smaller values of $\Delta $ indicate
 	better agreement with the exact ideal optimum.
 	
 	Our simulations show that bit-flip and phase-flip noise do not affect the
 	optimized \ac{QAOA} cost equally. For the \ac{FGA} instance considered here,
 	bit-flip noise consistently affects the application more severely. 
 	This
 	asymmetry is clearest in the layer-level model and in the native- $R_{ZZ} $ gate-level decomposition. In the CNOT-based decomposition, the additional gates required to implement each $ ZZ$ interaction introduce more noisy locations and allows for phase-flip errors to propagate, which makes phase-flip noise more significant as well. Nevertheless, bit-flip noise remains the dominant source of degradation.
 	
 	The effect remains visible when both types of errors are present at the same time. 
 	The residual cost is smallest near the noiseless point and increases as either  $p_X$  or $ p_Z$  is raised. 
 	The layer-level model gives the smallest overall degradation because noise is applied only once per QAOA layer, while the gate-level models show larger degradation due to noise after each elementary gate. The comparison between the CNOT and $R_{ZZ}$ decompositions further shows that the observed noise sensitivity depends not only on the physical error channel, but also on the circuit compilation.

 	The purpose of comparing different noise models
 	and gate decompositions is to investigate the
 	relative sensitivity to bit-flip and phase-flip
 	errors within each implementation, rather than
 	to compare their absolute performance, since
 	the models differ in the number and placement
 	of noisy operations.
 	
 	Our performance metric characterizes the best
 	optimized cost obtained from multiple random
 	initializations under a fixed optimization protocol,
 	while the spread across these initializations is
 	illustrated in the corresponding figures.
 	The penalty coefficients are also kept fixed
 	throughout the analysis, so the dependence of
 	the observed noise asymmetry on their values
 	remains to be investigated.
 	
 	Although our numerical investigation is limited to
 	a single six-qubit FGA instance, the mechanisms
 	underlying the observed noise asymmetry are not
 	specific to this problem. The commutation relations
 	between the error operators and the QAOA Hamiltonians,
 	together with their effects on the final measurements,
 	provide a physical explanation for the different
 	sensitivities to bit-flip and phase-flip noise.
 	
 	The persistence of the observed trend across the
 	considered circuit depths, noise models, and gate
 	decompositions further supports our expectation
 	that similar asymmetries may arise in other
 	Ising-based QAOA applications. 
 	Establishing the
 	extent of this generality will require further
 	investigations of different problem instances
 	and Hamiltonians.
 	
 	Finally, we used this application-level noise asymmetry to motivate asymmetric quantum error correction. 
 	Using the \ac{QPC} as an illustrative example, 
 	we provided strong evidence that the asymmetric codes can achieve nearly the same performance as a larger symmetric code, while using fewer physical qubits. 
 	In our analysis the $(5,3)$-\ac{QPC} performs as well as the $(5,5)$-\ac{QPC} using ten qubits less.
 	This suggests that error-correction resources should be assigned according to the error channel that is most damaging for the algorithmic objective, not only according to the physical error probability.
 	This illustrates that asymmetric \ac{QEC} codes can be useful even if the noise is symmetric.
 	

 	Although we expect this to be computationally expensive, future work could extend our approach to a full-fledged simulation of the quantum error correction circuit.
 	To avoid the computational overhead of the classical optimization loop of variational algorithms in the simulation, it seems promising to extend our approach to other algorithms relevant in the era of early fault-tolerance, like quantum walks~\cite{Venegas-Andraca2012, incudini2026practicaladvantagequadraticspeedup}.
 	Furthermore it seems promising to combine the analysis presented here with additional information on the noise obtained, for example, from calibration experiments like the one introduced in \cite{Wimmer2024}.
 	Future work might also investigate the advantage of asymmetric \ac{QEC} codes on real devices, also going to larger problem instances.
 	
 	\paragraph{Data availability}
 	The original contributions presented in this study are included in the supplementary material. Further inquiries can be directed to the corresponding author.
 	
 	\acknowledgments{ME and PS acknowledge funding by the Quantum Computing Initiative (QCI) of \ac{DLR} via project R-QIP.}



 \end{document}

%% file: plots/from_p_to_pL.pdf_tex
\begingroup%
  \makeatletter%
  \providecommand\color[2][]{%
    \errmessage{(Inkscape) Color is used for the text in Inkscape, but the package 'color.sty' is not loaded}%
    \renewcommand\color[2][]{}%
  }%
  \providecommand\transparent[1]{%
    \errmessage{(Inkscape) Transparency is used (non-zero) for the text in Inkscape, but the package 'transparent.sty' is not loaded}%
    \renewcommand\transparent[1]{}%
  }%
  \providecommand\rotatebox[2]{#2}%
  \newcommand*\fsize{\dimexpr\f@size pt\relax}%
  \newcommand*\lineheight[1]{\fontsize{\fsize}{#1\fsize}\selectfont}%
  \ifx\svgwidth\undefined%
    \setlength{\unitlength}{360bp}%
    \ifx\svgscale\undefined%
      \relax%
    \else%
      \setlength{\unitlength}{\unitlength * \real{\svgscale}}%
    \fi%
  \else%
    \setlength{\unitlength}{\svgwidth}%
  \fi%
  \global\let\svgwidth\undefined%
  \global\let\svgscale\undefined%
  \makeatother%
  \begin{picture}(1,1.01388887)%
    \lineheight{1}%
    \setlength\tabcolsep{0pt}%
    \put(0,0){\includegraphics[width=\unitlength,page=1]{from_p_to_pL.pdf}}%
    \put(0.24128619,0.89711425){\makebox(0,0)[lt]{\lineheight{1.25}\smash{\begin{tabular}[t]{l}Shor code\end{tabular}}}}%
    \put(0,0){\includegraphics[width=\unitlength,page=2]{from_p_to_pL.pdf}}%
    \put(0.24128619,0.82039117){\makebox(0,0)[lt]{\lineheight{1.25}\smash{\begin{tabular}[t]{l}First bisector\end{tabular}}}}%
    \put(0,0){\includegraphics[width=\unitlength,page=3]{from_p_to_pL.pdf}}%
    \put(0.24128619,0.74366808){\makebox(0,0)[lt]{\lineheight{1.25}\smash{\begin{tabular}[t]{l}$\Delta=C_{\mathrm{best}}-\lambda_{\min}$\end{tabular}}}}%
    \put(0,0){\includegraphics[width=\unitlength,page=4]{from_p_to_pL.pdf}}%
    \put(0.07926026,0.06966966){\makebox(0,0)[lt]{\lineheight{1.25}\smash{\begin{tabular}[t]{l}$0$\end{tabular}}}}%
    \put(0,0){\includegraphics[width=\unitlength,page=5]{from_p_to_pL.pdf}}%
    \put(0.45396521,0.06966966){\makebox(0,0)[lt]{\lineheight{1.25}\smash{\begin{tabular}[t]{l}$p$\end{tabular}}}}%
    \put(0,0){\includegraphics[width=\unitlength,page=6]{from_p_to_pL.pdf}}%
    \put(0.31679027,0.06966966){\makebox(0,0)[lt]{\lineheight{1.25}\smash{\begin{tabular}[t]{l}$p_L$\end{tabular}}}}%
    \put(0,0){\includegraphics[width=\unitlength,page=7]{from_p_to_pL.pdf}}%
    \put(0.06011285,0.0997261){\makebox(0,0)[lt]{\lineheight{1.25}\smash{\begin{tabular}[t]{l}0\end{tabular}}}}%
    \put(0,0){\includegraphics[width=\unitlength,page=8]{from_p_to_pL.pdf}}%
    \put(0.84781758,0.0997261){\makebox(0,0)[lt]{\lineheight{1.25}\smash{\begin{tabular}[t]{l}0\end{tabular}}}}%
    \put(0,0){\includegraphics[width=\unitlength,page=9]{from_p_to_pL.pdf}}%
    \put(0.84781761,0.42372348){\makebox(0,0)[lt]{\lineheight{1.25}\smash{\begin{tabular}[t]{l}$\Delta(p)$\end{tabular}}}}%
    \put(0,0){\includegraphics[width=\unitlength,page=10]{from_p_to_pL.pdf}}%
    \put(0.84781761,0.38602352){\makebox(0,0)[lt]{\lineheight{1.25}\smash{\begin{tabular}[t]{l}$\Delta(p_L)$\end{tabular}}}}%
    \put(0.35044966,0.01480338){\makebox(0,0)[lt]{\lineheight{1.25}\smash{\begin{tabular}[t]{l}error rate\end{tabular}}}}%
    \put(0.99175887,0.4508301){\rotatebox{90}{\makebox(0,0)[lt]{\lineheight{1.25}\smash{\begin{tabular}[t]{l}$\Delta$ (arb.units)\end{tabular}}}}}%
    \put(0,0){\includegraphics[width=\unitlength,page=11]{from_p_to_pL.pdf}}%
    \put(0.03520508,0.42210016){\rotatebox{90}{\makebox(0,0)[lt]{\lineheight{1.25}\smash{\begin{tabular}[t]{l}logical error rate\end{tabular}}}}}%
  \end{picture}%
\endgroup%

%% file: plots/QPC53.pdf_tex
\begingroup%
  \makeatletter%
  \providecommand\color[2][]{%
    \errmessage{(Inkscape) Color is used for the text in Inkscape, but the package 'color.sty' is not loaded}%
    \renewcommand\color[2][]{}%
  }%
  \providecommand\transparent[1]{%
    \errmessage{(Inkscape) Transparency is used (non-zero) for the text in Inkscape, but the package 'transparent.sty' is not loaded}%
    \renewcommand\transparent[1]{}%
  }%
  \providecommand\rotatebox[2]{#2}%
  \newcommand*\fsize{\dimexpr\f@size pt\relax}%
  \newcommand*\lineheight[1]{\fontsize{\fsize}{#1\fsize}\selectfont}%
  \ifx\svgwidth\undefined%
    \setlength{\unitlength}{450.99998474bp}%
    \ifx\svgscale\undefined%
      \relax%
    \else%
      \setlength{\unitlength}{\unitlength * \real{\svgscale}}%
    \fi%
  \else%
    \setlength{\unitlength}{\svgwidth}%
  \fi%
  \global\let\svgwidth\undefined%
  \global\let\svgscale\undefined%
  \makeatother%
  \begin{picture}(1,0.50332594)%
    \lineheight{1}%
    \setlength\tabcolsep{0pt}%
    \put(0,0){\includegraphics[width=\unitlength,page=1]{QPC53.pdf}}%
    \put(0.08936159,0.04680136){\makebox(0,0)[lt]{\lineheight{1.25}\smash{\begin{tabular}[t]{l}0.000\end{tabular}}}}%
    \put(0,0){\includegraphics[width=\unitlength,page=2]{QPC53.pdf}}%
    \put(0.24715748,0.04680136){\makebox(0,0)[lt]{\lineheight{1.25}\smash{\begin{tabular}[t]{l}0.005\end{tabular}}}}%
    \put(0,0){\includegraphics[width=\unitlength,page=3]{QPC53.pdf}}%
    \put(0.40495334,0.04680136){\makebox(0,0)[lt]{\lineheight{1.25}\smash{\begin{tabular}[t]{l}0.010\end{tabular}}}}%
    \put(0,0){\includegraphics[width=\unitlength,page=4]{QPC53.pdf}}%
    \put(0.5627492,0.04680136){\makebox(0,0)[lt]{\lineheight{1.25}\smash{\begin{tabular}[t]{l}0.015\end{tabular}}}}%
    \put(0,0){\includegraphics[width=\unitlength,page=5]{QPC53.pdf}}%
    \put(0.72054506,0.04680136){\makebox(0,0)[lt]{\lineheight{1.25}\smash{\begin{tabular}[t]{l}0.020\end{tabular}}}}%
    \put(0,0){\includegraphics[width=\unitlength,page=6]{QPC53.pdf}}%
    \put(0.03467988,0.08889794){\makebox(0,0)[lt]{\lineheight{1.25}\smash{\begin{tabular}[t]{l}0.00\end{tabular}}}}%
    \put(0,0){\includegraphics[width=\unitlength,page=7]{QPC53.pdf}}%
    \put(0.03467988,0.20051948){\makebox(0,0)[lt]{\lineheight{1.25}\smash{\begin{tabular}[t]{l}0.05\end{tabular}}}}%
    \put(0,0){\includegraphics[width=\unitlength,page=8]{QPC53.pdf}}%
    \put(0.03467988,0.31214099){\makebox(0,0)[lt]{\lineheight{1.25}\smash{\begin{tabular}[t]{l}0.10\end{tabular}}}}%
    \put(0,0){\includegraphics[width=\unitlength,page=9]{QPC53.pdf}}%
    \put(0.03467988,0.42376252){\makebox(0,0)[lt]{\lineheight{1.25}\smash{\begin{tabular}[t]{l}0.15\end{tabular}}}}%
    \put(0,0){\includegraphics[width=\unitlength,page=10]{QPC53.pdf}}%
    \put(0.40476043,0.00879555){\makebox(0,0)[lt]{\lineheight{1.25}\smash{\begin{tabular}[t]{l}$p_x=p_z$\end{tabular}}}}%
    \put(0.01810161,0.15){\rotatebox{90}{\makebox(0,0)[lt]{\lineheight{1.25}\smash{\begin{tabular}[t]{l}$\Delta = C_{\mathrm{best}} - \lambda_{\mathrm{min}}$\end{tabular}}}}}%
    \put(0,0){\includegraphics[width=\unitlength,page=11]{QPC53.pdf}}%
    \put(0.87960091,0.32141443){\makebox(0,0)[lt]{\lineheight{1.25}\smash{\begin{tabular}[t]{l}No QEC\end{tabular}}}}%
    \put(0,0){\includegraphics[width=\unitlength,page=12]{QPC53.pdf}}%
    \put(0.87960091,0.27063837){\makebox(0,0)[lt]{\lineheight{1.25}\smash{\begin{tabular}[t]{l}(5,3)\end{tabular}}}}%
    \put(0,0){\includegraphics[width=\unitlength,page=13]{QPC53.pdf}}%
    \put(0.87960091,0.21986232){\makebox(0,0)[lt]{\lineheight{1.25}\smash{\begin{tabular}[t]{l}(3,3)\end{tabular}}}}%
    \put(0,0){\includegraphics[width=\unitlength,page=14]{QPC53.pdf}}%
    \put(0.87960091,0.16908627){\makebox(0,0)[lt]{\lineheight{1.25}\smash{\begin{tabular}[t]{l}(5,5)\end{tabular}}}}%
  \end{picture}%
\endgroup%